\documentclass[column]{aa} 

\usepackage{natbib,twoopt}
\usepackage[utf8]{inputenc}

\usepackage[breaklinks=true]{hyperref} 
\bibpunct{(}{)}{;}{a}{}{,}             
\makeatletter
  \newcommandtwoopt{\citeads}[3][][]{\href{http://adsabs.harvard.edu/abs/#3}%
    {\def\hyper@linkstart##1##2{}%
     \let\hyper@linkend\@empty\citealp[#1][#2]{#3}}}
  \newcommandtwoopt{\citepads}[3][][]{\href{http://adsabs.harvard.edu/abs/#3}%
    {\def\hyper@linkstart##1##2{}%
     \let\hyper@linkend\@empty\citep[#1][#2]{#3}}}
  \newcommandtwoopt{\citetads}[3][][]{\href{http://adsabs.harvard.edu/abs/#3}%
    {\def\hyper@linkstart##1##2{}%
     \let\hyper@linkend\@empty\citet[#1][#2]{#3}}}
  \newcommandtwoopt{\citeyearads}[3][][]%
    {\href{http://adsabs.harvard.edu/abs/#3}
    {\def\hyper@linkstart##1##2{}%
     \let\hyper@linkend\@empty\citeyear[#1][#2]{#3}}}
\makeatother

\usepackage{mathtools, amsmath}
\usepackage{nccmath}
\usepackage{float}
\usepackage{graphicx}
\usepackage{hyperref}
\usepackage{placeins}
\usepackage{ulem}
\usepackage{hhline}
\usepackage{svg}

\newcommand{\linkorcid}[1]{\href{https://orcid.org/#1}{\includegraphics[width=8pt]{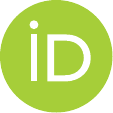}}}

\usepackage{xcolor}
\hypersetup{
    colorlinks,
    linkcolor={red!50!black},
    citecolor={blue!50!black},
    urlcolor={blue!80!black}
}
\usepackage{natbib}

\newcommand{\hmpc }{$h^{-1}$Mpc}

\newcommand{\sn}{SN~Ia}
\newcommand{\sns}{SNe~Ia}
\newcommand{\kms}{km s$^{-1}$}
\newcommand{\kmsmpc}{km~s$^{-1}$~Mpc$^{-1}$}

\usepackage[switch]{lineno} 

\begin{document} 

\linenumbers
\renewcommand\makeLineNumber{}

   \title{ZTF SN~Ia DR2: Peculiar velocities' impact on the Hubble diagram}


   \author{
        B. Carreres\inst{\ref{CPPM}, \ref{Duke}}\fnmsep\thanks{Corresponding author \email{bastien.carreres@duke.edu}}\linkorcid{0000-0002-7234-844X} 
        \and D. Rosselli \inst{\ref{CPPM}}\fnmsep\thanks{Corresponding author \email{rosselli@cppm.in2p3.fr}}\linkorcid{0000-0001-6839-1421}
        \and  J.~E.~Bautista\inst{\ref{CPPM}}\linkorcid{0000-0002-9885-3989}
        \and F.~Feinstein\inst{\ref{CPPM}}\linkorcid{0000-0001-5548-3466}
        \and D.~Fouchez\inst{\ref{CPPM}}\linkorcid{0000-0002-7496-3796 }
        \and B.~Racine\inst{\ref{CPPM}}\linkorcid{0000-0001-8861-3052}
        \and C.~Ravoux\inst{\ref{CPPM}, \ref{LPC}}\linkorcid{0000-0002-3500-6635}
        \and B.~Sanchez\inst{\ref{CPPM}}
        \and G.~Dimitriadis\inst{\ref{Trinity}}\linkorcid{0000-0001-9494-179X}
        \and A.~Goobar\inst{\ref{OKC}}
        \and J.~Johansson\inst{\ref{OKC}}\linkorcid{0000-0001-5975-290X}
        \and J.~Nordin\inst{\ref{IfPBerlin}}\linkorcid{0000-0001-8342-6274}
        \and M.~Rigault\inst{\ref{IP2I}}\linkorcid{0000-0002-8121-2560}
        \and M.~Smith\inst{\ref{Lancaster}}\linkorcid{0000-0002-3321-1432}
        \and M.~Amenouche\inst{\ref{NRCC}}\linkorcid{0009-0006-7454-3579}
        \and M.~Aubert\inst{\ref{LPC}}
        \and C.~Barjou-Delayre\inst{\ref{LPC}}
        \and U.~Burgaz\inst{\ref{Trinity}}\linkorcid{0000-0003-0126-3999}
        \and W. D’Arcy Kenworthy\inst{\ref{OKC}}\linkorcid{0000-0002-5153-5983}
        \and T.~De Jaeger\inst{\ref{LPNHE}}
        \and S. Dhawan\inst{\ref{Cambridge}}\linkorcid{0000-0002-2376-6979}
        \and L.~Galbany\inst{\ref{ISS}}
        \and M.~Ginolin\inst{\ref{IP2I}}\linkorcid{0009-0004-5311-9301}
        \and D.~Kuhn\inst{\ref{LPNHE}}\linkorcid{0009-0005-8110-397X}
        \and M. Kowalski\inst{\ref{IfPBerlin}, \ref{DESY}}\linkorcid{0000-0001-8594-8666}
        \and T.~E.~Müller-Bravo\inst{\ref{ISS}}\linkorcid{0000-0003-3939-7167}
        \and P.~E.~Nugent\inst{\ref{lbnl}, \ref{berkeley}}\linkorcid{0000-0002-3389-0586}
        \and B.~Popovic\inst{\ref{IP2I}}
        \and P.~Rosnet\inst{\ref{LPC}}
        \and F.~Ruppin\inst{\ref{IP2I}}
        \and J.~Sollerman\inst{\ref{OKC}}\linkorcid{0000-0003-1546-6615}
        \and J.~H.~Terwel\inst{\ref{Trinity}, \ref{NOT}} \linkorcid{0000-0001-9834-3439}
        \and A.~Townsend\inst{\ref{IfPBerlin}}\linkorcid{0000-0001-6343-3362}
        \and S.~L.~Groom\inst{\ref{IPAC}}
        \and S.~R.~Kulkarni\inst{\ref{249Caltech}}
        \and J.~Purdum\inst{\ref{Caltech}}
        \and B.~Rusholme\inst{\ref{IPAC}}
         \and N.~Sravan\inst{\ref{Drexel}}
          }
         \institute{
             Aix Marseille Université, CNRS/IN2P3, CPPM, Marseille, France\label{CPPM}
        \and Department of Physics, Duke University, Durham, NC 27708, USA\label{Duke}
        \and Université Clermont-Auvergne, CNRS, LPCA, 63000 Clermont-Ferrand, France\label{LPC}
        \and School of Physics, Trinity College Dublin, College Green, Dublin 2, Ireland\label{Trinity}
        \and Oskar Klein Centre, Department of Astronomy, Stockholm University, SE-10691 Stockholm, Sweden\label{OKC}
        \and Institut für Physik, Humboldt Universität zu Berlin, Newtonstr 15, 12101 Berlin\label{IfPBerlin}
        \and Univ Lyon, Univ Claude Bernard Lyon 1, CNRS, IP2I Lyon/IN2P3, UMR 5822, F-69622, Villeurbanne, France\label{IP2I}
        \and Department of Physics, Lancaster University, Lancs LA1 4YB, UK\label{Lancaster}
        \and National Research Council of Canada, Herzberg Astronomy \& Astrophysics Research Centre, 5071 West Saanich Road, Victoria, BC V9E 2E7, Canada\label{NRCC}
        \and Sorbonne/Paris Cité Universités, LPNHE, CNRS/IN2P3, LPNHE, 75005, Paris, France\label{LPNHE}
        \and Institute of Astronomy and Kavli Institute for Cosmology, University of Cambridge, Madingley Road, Cambridge CB3 0HA, UK\label{Cambridge}
        \and Institute of Space Sciences (ICE, CSIC), Campus UAB, Carrer de Can Magrans, s/n, E-08193, Barcelona, Spain\label{ISS}
        \and Deutsches Elektronen Synchrotron DESY, Platanenallee 6, 15738, Zeuthen\label{DESY}
        \and Lawrence Berkeley National Laboratory, 1 Cyclotron Road MS 50B-4206, Berkeley, CA, 94720, USA\label{lbnl}
        \and Department of Astronomy, University of California, Berkeley, 501 Campbell Hall, Berkeley, CA 94720, USA\label{berkeley}
        \and Nordic Optical Telescope, Rambla José Ana Fernández Pérez 7, ES-38711 Breña Baja, Spain\label{NOT}
         \and IPAC, California Institute of Technology, 1200 E. California Blvd, Pasadena, CA 91125, USA\label{IPAC}
         \and 249-17 Caltech, Pasadena, CA 91125, USA\label{249Caltech}
         \and Caltech Optical Observatories, California Institute of Technology, Pasadena, CA 91125, USA\label{Caltech}
        \and Department of Physics, Drexel University, Philadelphia, PA 19104, USA\label{Drexel}
        }

   \date{}

  \abstract
    {Type Ia supernovae (\sns)\ are used to determine the distance-redshift relation and build the Hubble diagram. Neglecting their host-galaxy peculiar velocities (PVs) may bias the measurement of cosmological parameters. The smaller the redshift, the larger the effect is. We used realistic simulations of \sns\ observed by the \textit{Zwicky} Transient Facility (ZTF) to investigate the effect of different methods of taking PVs into account. We studied the impact of neglecting galaxy PVs and their correlations in an analysis of the \sns\ Hubble diagram. We find that it is necessary to use the PV full covariance matrix computed from the velocity power spectrum to take the sample variance into account. Considering the results we have obtained using simulations, we determine the PV systematic effects in the context of the ZTF SN~Ia DR2 sample. We determine the PV impact on the intercept of the Hubble diagram, $a_B$, which is directly linked to the measurement of $H_0$. We show that not taking into account PVs and their correlations results in a shift in the $H_0$ value of about $1.0$~\kmsmpc and a slight underestimation of the $H_0$ error bar.
    } 
    

    \titlerunning{ZTF SN~Ia DR2: Peculiar velocities' impact on the Hubble diagram}
    \authorrunning{B. Carreres, D. Rosselli et al. }
   \maketitle
%

\section{Introduction}

Type Ia supernovae (\sns) are well-known standardizable candles that have been widely used to provide precise estimates of the luminosity distances to galaxies, as an input of the  Hubble diagram. \sns\ are an essential probe with which to measure cosmological parameters such as the dark matter density, the dark energy density, its equation-of-state parameter, $w$ \citep{betouleImprovedCosmologicalConstraints2014,broutPantheonAnalysisCosmological2022}, and the Hubble constant, $H_0$\citep{riessComprehensiveMeasurementLocal2022, scolnic2023cats}.

Several surveys have been designed (e.g., LSST, \citealt{LSST_2009_sciencebook}; \textit{Euclid}, \citealt{laureijs2011euclid}; and DESI, \citealt{desicollaboration2016desi}) to collect powerful datasets for cosmology. In particular, the \textit{Zwicky} Transient Facility (ZTF) was designed to observe several thousands of \sn\ light curves, by far the largest low-redshift sample ($z < 0.1$) to date. By combining the data from all these surveys, cosmological parameters, such as $H_0$, $w$, and the matter density parameter, $\Omega_m$, will be measured with a precision never reached before, allowing us to put strong constraints on the $\Lambda$CDM model. To achieve this goal, we need to improve our understanding of different sources of systematic uncertainties that were negligible in the past compared to statistical uncertainties. In this article, we explore the effect of the \sn\ host-galaxy peculiar velocities (PVs) on the determination of $H_0$ with the ZTF sample.

Peculiar velocities refer to the motion of galaxies with respect to a comoving frame in an expanding Universe. They originate from the gravitational infall of matter toward overdense regions.
As has been demonstrated in previous works such as \citealt{huiCorrelatedFluctuationsLuminosity2006,davisEffectPeculiarVelocities2011,petersonPantheonAnalysisEvaluating2022}, the SN Ia host galaxy PVs have the most significant impact on cosmological parameter inference when using low-redshift {\sns}. Depending on the sign of the PV projection on the observer’s line of sight, it will produce either a blue or a red Doppler shift ($z_p$) of the observed redshift ($z_\mathrm{obs}$) with respect to the cosmological redshift ($z_\mathrm{cos}$):
\begin{equation}
    1+z_\mathrm{obs}=(1+z_\mathrm{cos})(1+z_p).
    \label{eq:zcomp}
\end{equation}

The PV distribution of the host galaxies will affect the redshift-distance relation in two ways. Firstly, PVs  will increase the scatter in 
the Hubble diagram residuals, since 
the observed redshifts are used to compute the predicted
distance modulus, $\mu_\mathrm{th}(z_{\mathrm{obs}})$, for a given cosmology and $\mu_\mathrm{th}(z_{\mathrm{obs}})$ is directly related to the Hubble diagram residuals. We can consider that, to the first order, $z_p = v_p/c$, where $v_p$ is the galaxy PV line-of-sight component and $c$ is the speed of light. The typical $v_p $ is on the order of several hundred {\kms}. At low redshift, it corresponds to a magnitude scatter comparable to the \sn\ intrinsic scatter (see \citealt{davisEffectPeculiarVelocities2011, carreresGrowthrateMeasurementTypeIa2023} and references therein for a detailed discussion).

Secondly, PVs are correlated spatially due to large-scale bulk motion \citep[see][]{peebles_LSS}. The majority of previous \sn\ analyses consider PVs to be independent and randomly distributed. However, there are non-negligible correlations between galaxy velocities in different sky positions. These correlations arise because PVs are the result of galaxy motion within the same large-scale structure gravitational potential. Neglecting the PV spatial correlation when fitting the Hubble diagram may bias cosmological parameter measurements \citep{davisEffectPeculiarVelocities2011}, especially when analyzing low-redshift \sn\ samples. 

When both the distance and the redshift to a galaxy are available, one can derive an estimate of its PV. Studying the correlations of such velocities allows us to constrain the cosmic growth rate of structures, $f\sigma_8$. This parameter expresses the growth of the cosmic structure throughout the history of the Universe. Various methods have been developed to perform this measurement (see \citealt{johnson6dFGalaxyVelocity2014,howlett2MTFVIMeasuring2017,saidJointAnalysis6dFGS2020,turnerLocalMeasurementGrowth2022,laiUsingPeculiarVelocity2023,carreresGrowthrateMeasurementTypeIa2023}).

The Hubble constant, $H_0$, could be measured using \sns\ with distance anchors to calibrate the absolute magnitude, $M_B$. The usual anchors are nearby galaxy distances measured thanks to Cepheids \citep{riessComprehensiveMeasurementLocal2022} or the tip of the red giant branch (TRGB, \citealt{dhawanUniformTypeIa2022, scolnic2023cats}). This allows one to obtain the absolute magnitude, $M_B$, of the nearby {\sns}. The \sns\ that reside in the Hubble flow are used to determine the intercept of the Hubble diagram, $a_B$. The intercept is linked to $H_0$ through the relation 
\begin{equation}\label{H0-ab}
    \log{H_0} = \frac{M_B + 5a_B + 25}{5}.
\end{equation}

In recent analysis \citep{broutPantheonAnalysisCosmological2022, riessComprehensiveMeasurementLocal2022, Darcy_h02rungs_2022}, PVs have been corrected using PV field reconstruction methods \citep{lavaux2MGalaxyRedshift2011, carrickCosmologicalParametersComparison2015, grazianiPeculiarVelocityField2019, Lilow2021}. The first attempts at applying these methods to correct PVs to \sns\ in the Hubble diagram have shown improvements in the residuals \citep{petersonPantheonAnalysisEvaluating2022, carrPantheonAnalysisImproving2022}. However, these methods require a spectroscopic survey of the underlying large-scale structure. Moreover, the current literature lacks testing of these methods on simulations, which would provides a better understanding of the bias and systematics that could arise from their use in the Hubble diagram.


In this work, we study the impact of galaxy PVs and their correlations on the Hubble diagram fit in the context of the  \sns\ sample of the second ZTF Data Release from the ZTF Type Ia Supernovae \& Cosmology
science working group (ZTF SN~Ia DR2). We make use of both realistic simulations and real data to study the systematic impact of PVs in the Hubble diagram. We determine the effect of PVs on the intercept of the Hubble diagram $a_B$. This work will be useful for future cosmological analyses that may use the \sns\ coming from ZTF survey.

The paper is organized as follows. In Sect. \ref{Datasec}, we describe the data and the simulations we have used for this work. In Sect. \ref{Methodsec}, we present the method used to fit the Hubble diagram and to take into account PVs. In Sect. \ref{simulationresults}, we describe our findings when we analyze the realistic \sn\ simulations. In
Sect. \ref{sec:IntercepthubbleDR2}, we present our main result on the impact of PVs on the intercept of the Hubble diagram for the second data-release of the ZTF Type Ia Supernova survey. Finally, we state our conclusions in Sect.~\ref{sec:conclusion}.

\section{Data}\label{Datasec}

\subsection{The ZTF SN~Ia DR2 sample}\label{sec:dr2}
The ZTF (\citealt{bellmZwickyTransientFacility2019,graham_ztf_2019,masci_2019_ztf,dekanyZwickyTransientFacility2020}) is an optical survey that started operating in $2018$. It covers two thirds of the sky from a declination of $-30^{\circ}$ to a declination of $+90^{\circ}$. Thanks to its 47 deg$^2$ field of view, it observes in the $g$, $r$, and $i$ bands with a typical magnitude depth of 20 and a cadence ranging from several visits per day to several visits per month. The detected transient sources are followed up spectroscopically for precise classification. The reduction of spectra and classification are performed by the ZTF SED machine \citep{blagorodnovaSEDMachineRobotic2018,rigaultFullyAutomatedIntegral2019, KimNewModSEDm}.

The \sn\ sample used in this work is the ZTF SN~Ia DR2 sample (\citealt{rigaultZTFoverview}, Smith et al. in prep.).  It is a sample of \sn\ data coming from the ZTF survey between March 2018 and December 2020. It consists of $3628$ \sn\ light curves, with host galaxy properties (e.g., stellar mass, g-z). All the \sns\ have been spectroscopically classified and have spectroscopic redshifts that come from the host galaxy ($\sim 70~\%$), mostly given by the DESI MOST host program \citep{Maayane2024}, or from the \sn\ spectrum ($\sim 30~\%$). The host galaxy redshifts have an uncertainty of about $10^{-5}$, while the \sn\ spectrum redshifts have an uncertainty of about $10^{-3}$ (see \citealt{smith} for details about the \sn\ redshifts and the method of matching the \sn\ with its host galaxy). The light curves were fitted using the SALT2 model \citep{guySALT2UsingDistant2007,guySupernovaLegacySurvey2010} and the release contains the resulting parameters: the time of the peak magnitude, $t_0$, the stretch, $x_1$, the color, $c$, and the peak magnitude in the Bessel-B band at rest frame, $m_B$.
The DR2 \sns\ are volume-limited up to $z \sim 0.06$ (\citealt{carreresGrowthrateMeasurementTypeIa2023}, Amenouche et al. in prep.). This constitutes the largest low-redshift \sns\ sample to date.

In this work, we applied to the \sn\ light curves the quality cuts given in Table~1 of \cite{rigaultZTFoverview}. We required at least five observations with a signal-to-noise ratio above five to ensure a good sampling of the light curves. We selected the light curves with best-fit values for stretch, $|x_1|<3$, and color, $-0.2 < c < 0.8$. We excluded the \sns\ with too-large estimated uncertainties on those parameters by requiring that $\sigma_{x_1} < 1$, $\sigma_c<0.1$ and  $\sigma_{t_0}<1$. We discarded poor fit models by requiring a $p$-value above $10^{-7}$\ .\footnote{For a resulting $\chi^2$ from a fit with $k$ degrees of freedom, it is equivalent to require $\int_{\chi^2}^{+\infty}f_{\chi^2, k}(x)dx > 10^{-7}$, where $f_{\chi^2,k}(x)$ is the $\chi^2$ distribution for $k$ degrees of freedom.} Lastly, we selected \sns\ within the redshift range of $0.023 < z_\mathrm{obs} < 0.06$. The low-redshift bound exludes \sn\ velocities correlated within our local flow, while the high-redshift bound ensures that our sample is volume-limited. After these cuts, we were left with 904 \sns. We accessed the ZTF SN~Ia DR2 data using the python library \texttt{ztfidr}.\footnote{\url{https://github.com/MickaelRigault/ztfidr}}

\subsection{Simulations}

To test our framework, we used simulations of ZTF \sns\ including large-scale structure, as was performed in \citet{carreresGrowthrateMeasurementTypeIa2023}. These simulations were produced using the \texttt{snsim}\footnote{\url{https://github.com/bastiencarreres/snsim}} python library. This software allows one to simulate a realistic \sns\ survey on top of an N-body simulation. In this work, we used the OuterRim N-body simulation \citep{heitmannOuterRimSimulation2019}, which is a box of $(3 \ \mathrm{Gpc})^3$ at $z=0$. From this box, we created 27 sub-boxes. In order to minimize any possible correlation between them, the sub-boxes were built so that they did not overlap. For each sub-box, the observer was placed at the center and the enclosed volume was equivalent to a redshift limit of $z\sim 0.17$. In Appendix~\ref{app:subboxescorr}, we discuss the correlation between the different sub-boxes. We found that the correlations between the host galaxies below the redshift cut of $z\sim0.06$ in different sub-boxes are one order of magnitude below the ones between two host galaxies that are in the same mock, and thus can safely be neglected.

To be consistent, the fiducial cosmology used in our simulations matches the one used in OuterRim: a flat-$\Lambda$CDM model with parameters provided in Table~\ref{tab:siminput}.
\begin{table}
    \centering
    \caption{Cosmological parameters used in the simulation.}
        \begin{tabular}{ccccccc} 
            \hline 
            \hline \\[-3ex]
            $H_0$ & $\omega_{\rm cdm}$ &  $\omega_{\rm b}$ & $n_s$ & $\sigma_8$ & $f$ \\ 
            \hline \\[-1.8ex] 
            $71.$ & $0.1109$ & $0.02258$ & $0.963$ & $0.800 $ & $0.478$
        \end{tabular} 
    \tablefoot{$H_0$ is given in \kmsmpc. $f\sigma_8$ is given at $z=0$.}
    \label{tab:siminput}
\end{table} 
The number of \sn\ events to be generated was computed using the latest estimate of the rate measured from ZTF data, $r_\mathrm{P20} = 2.35 \times 10^{-5}$ Mpc$^{-3}$ yr$^{-1}$ \citep{perleyZwickyTransientFacility2020}. This rate had to be rescaled to our fiducial cosmology as $r = r_\mathrm{P20} \left(h/0.7\right)^3$ with $h= H_0 / (100 \mathrm{km.s}^{-1}.\mathrm{Mpc}^{-1})$. To get more statistics for the expected number of \sns\ per Mpc$^{-3}$, we multiplied this rate by the expected 6-year duration of the full ZTF survey.
Each \sn\ was randomly assigned to a halo from the N-body simulation and the halo velocity was used to compute the observed redshift. We computed the angular positions of each halo relative to the observer and kept only those lying within the ZTF angular footprint. We randomly drew the SALT2 model parameters of each \sn: the stretch, $x_1$, the color, $c$, and the apparent peak magnitude in the Bessel-$B$ band, $m_B$. The stretch parameter was modeled using the redshift-dependent two-Gaussian mixture described in \citet{nicolasRedshiftEvolutionUnderlying2021}, while the color parameter distribution follows the asymmetric model given in \citet[Table~1]{scolnicMeasuringTypeIa2016} for low $z$.

The apparent peak magnitude was computed starting from the Tripp formula \citep{trippTwoparameterLuminosityCorrection1998}, which gives the absolute magnitude, $M_{B,i}^*$, in the rest-frame Bessel-$B$ band for \sn\ $i$, defined as 
\begin{equation}
    M_{B,i}^* = M_B - \alpha x_{1,i} + \beta c_i + \sigma_{M,i} .
\end{equation}

The $\alpha$, $\beta$, and $M_B$ parameters are common to all the \sns. These parameters were fixed to best-fit values from \citet{betouleImprovedCosmologicalConstraints2014}: $\alpha = 0.14$, $\beta = 3.1$, and $M_B = -19.05 + 5\log(h/0.7)$, where $M_B$ is rescaled to the fiducial cosmology.
The intrinsic scatter, $\sigma_{M}$, was drawn from a normal distribution with a dispersion equal to $0.12$, close to what is observed \citep[see][Table ~9]{betouleImprovedCosmologicalConstraints2014}. 
Finally, the apparent peak magnitude for each \sn\ was computed using
\begin{equation}
    m_{B,i} =  M_{B,i}^* + \mu_\mathrm{obs},
\end{equation}
where $\mu_\mathrm{obs}$ is the observed distance modulus, where the relativistic beaming due to the PV effect \citep{davisEffectPeculiarVelocities2011} is taken into account and is given by
\begin{equation}
    \mu_\mathrm{obs} = \mu_\mathrm{cos} + 10 \log(1+z_{p}) .
\end{equation}

$\mu_\mathrm{cos}$ is the distance modulus computed for a given cosmological redshift in a given cosmological model as
\begin{equation}\label{cosmodistmodule}
    \mu_\mathrm{cos} = 5\log\left(\frac{d_L(z_\mathrm{cos})}{10 \ \mathrm{pc}}\right),
\end{equation}
where $d_L(z)$ is the luminosity distance, 
\begin{equation}
    d_L(z) = (1 + z) r(z),
\end{equation}
and $r(z)$  is the comoving distance.

We generated the \sns\ with the SALT2 model input parameters to compute the true light curve for each \sn. We made use of the \texttt{SNCOSMO} package\footnote{\url{https://sncosmo.readthedocs.io/}} \citep{barbarySNCosmo2023} to model the light curves.
Then we used the ZTF observation logs to obtain realistic observations from light-curve models, as is described in the following: the dates of observations and the filters used at the position of the \sn\ allow one to get a realistic sampling of the light curve, while the limiting magnitude at $5\sigma$ above the sky background, the CCD gain, and the zero point of each observation are used to get realistic fluxes and error bars. The error bars were drawn from a normal distribution, the scatter of which is given by Eq.~14 in \cite{carreresGrowthrateMeasurementTypeIa2023}. 

\begin{figure}
    \centering
    \includegraphics[width=\columnwidth]{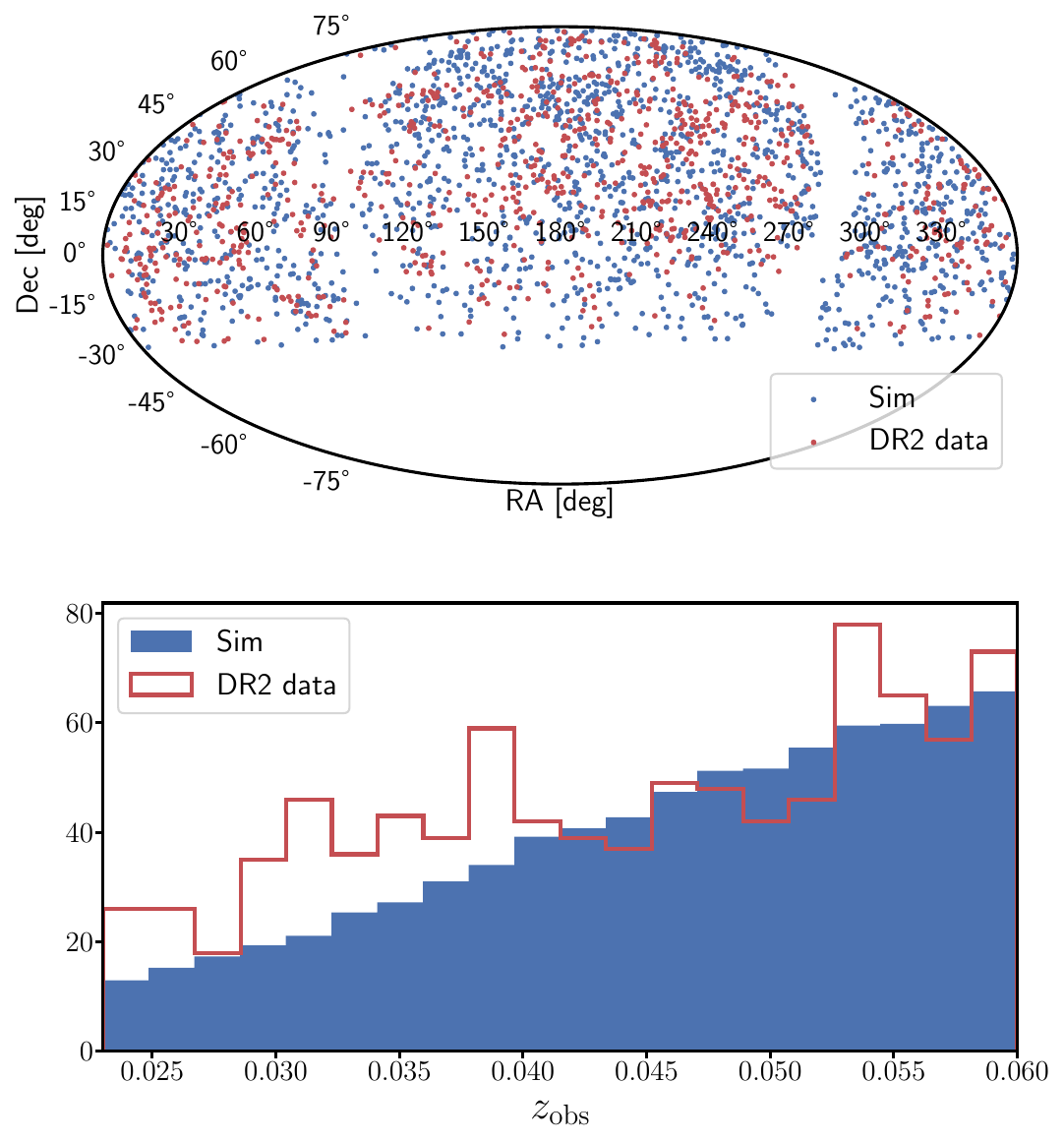}
    \caption{Comparison of the angular and redshift distributions of the SNe Ia from the simulation and from the ZTF SN~Ia DR2 data. \textit{Top panel}: Angular positions of simulated \sns\ from one mock (in blue) and from the DR2 data (in red).
    \textit{Bottom panel}: Redshift distribution averaged over the 27 mocks and rescaled for the difference in the duration of the survey (in blue) compared to the DR2 redshift distribution (in red).}
    \label{fig:simu_DR2}
\end{figure}

We applied to the simulated light curves the selections described in Sect.~2.5 of \citet{carreresGrowthrateMeasurementTypeIa2023}: we discarded the light curves with fewer than two epochs with fluxes $S/N > 5 $ and we applied the spectroscopic selection function of the ZTF Bright Transient Survey (BTS), the main survey used for the spectral classification of ZTF \sns\ \citep{perleyZwickyTransientFacility2020}.
After the selection, we fit the light curves using the SALT2 model to recover $m_b$, $x_1$, and $c$. We applied to the simulations the same quality cuts as for the real data (Sect.~\ref{sec:dr2}). After the quality cuts, the average number of \sns\ across each realization is $\langle N\rangle \sim 1672$ for the 6-year span of the full ZTF survey in the redshift range $0.023 < z < 0.06$. This number is lower than what could be found in the data since the ZTF SN~Ia DR2 span of $\sim 2.8$ years and has found $\sim 900$ \sns. We plot the comparison between simulations and DR2 data angular positions and redshifts distributions in Fig.~\ref{fig:simu_DR2}, in which the redshift distribution has been rescaled for the difference of duration between data and the simulations. The discrepancy between the number of \sns\ in a simulation and DR2 data could come from simulation inputs, such as a too-low value for the \sns\ rate, but also from the fact that the simulation spectroscopic selection function is only based on the BTS one, despite a small part of the DR2 data being classified using other spectroscopic programs (see \citealt{rigaultZTFoverview} for a full description).

The \sns\ observed redshifts and magnitudes are affected by correlated PVs. Hence, we used the 27 realizations to test various methods to take into account PVs when fitting the Hubble diagram, as is shown in the next sections.

\section{Method}\label{Methodsec}

We fit the Hubble diagram with a classical likelihood method that we describe in Sect.~\ref{Methodsec:HD}. However, we tested three ways of dealing with the PV error contribution: not accounting for PVs, including a diagonal error term, and using a full PV covariance matrix. These three approaches are described in Sect.~\ref{Methodsec:PV}.

\subsection{Hubble diagram}\label{Methodsec:HD}
Before fitting the Hubble diagram, we fit the \sn\ light curves using the SALT2 model. After the \sn\ light-curve fit, the distance modulus of each \sn\ $\mu_i$ was computed using the Tripp relation \citep{trippTwoparameterLuminosityCorrection1998} as
\begin{equation}
    \mu_i = m_{B,i} + \alpha x_{1,i} - \beta c_i - M_0,
    \label{eq:mu}
\end{equation}
where $\alpha$, $\beta$, and $M_0$ are free parameters of the Hubble diagram fit. 
The corresponding fit error bar for each \sn\ is
\begin{align}\label{eq:salterror}
    \sigma_{\mathrm{SALT},i}^2 = &\sigma^2_{m_B,i}+ \sigma_{x_1, i}^2 + \sigma_{c,i}^2 + 2\alpha \mathrm{COV}(m_B, x_1)\\
    &- 2\beta\mathrm{COV}(m_B,c) - 2\alpha\beta\mathrm{COV}(x_1, c) \nonumber .
\end{align}

To fit the Hubble diagram, we maximized the likelihood given by
\begin{equation}\label{eq:HD_likelihood}
    \mathcal{L}(\alpha, \beta, M_0) = (2\pi)^{-\frac{N}{2}} |C|^{-\frac{1}{2}}\exp\left[-\frac{1}{2}\mathbf{\Delta\mu}^T{C}^{-1}\mathbf{\Delta\mu}\right],
\end{equation}
where $C$ is a covariance matrix and the elements of $\mathbf{\Delta\mu}$ are 
\begin{equation}
    \Delta\mu_i = \mu_i - \mu_\mathrm{th}(z_{\mathrm{obs},i }),
\end{equation}
with $\mu_\mathrm{th}(z)$ the theoretical distance modulus (Eq.~\ref{cosmodistmodule}) computed for a given cosmology. Since ZTF is a low-redshift survey and cannot alone constrain cosmological parameters, we fixed them to a fiducial value. In our work on simulations (Sect.~\ref{simulationresults}) we used the input cosmology of the simulation as the fiducial one and in our work on data (Sect.~\ref{Datasec}) we used the \textit{Planck}18 results \citep{planckcollaborationPlanck2018Results2020} as fiducial cosmology.
It is worth noting that in this fit, $M_0$ and $H_0$ were degenerated. This necessitated that we choose a fiducial value, $H_{0,\mathrm{fid}}$, such that the $M_0$ parameter is expressed as a combination of $H_0$ and $M_B$ as
\begin{equation}
    M_0 = M_B - 5\log\left(\frac{H_0}{H_{0, \mathrm{fid}}}\right).
\end{equation}
The fiducial value of $H_0$ was taken to be equal to the simulation input, $H_{0, \mathrm{fid}} = 71$ \kmsmpc.

As is stated in Sect.~\ref{sec:dr2}, the ZTF SN~Ia DR2 redshifts were measured with varying accuracy. We took the redshift uncertainties into account by adding the following error term to the diagonal of the covariance matrix:
\begin{equation}\label{eq:rederror}
    \sigma_{\mu - z}^2 = \left(\frac{\partial \mu}{\partial z}\right)^2 \sigma_z^2 \simeq \left(\frac{5}{\ln10}\right)^2 \left(1 - z + \frac{1}{z}\right)^2 \sigma_z^2,
\end{equation} 
where we made an approximation at the first order with respect to $z$.

Without taking into account the PVs, the covariance matrix is 
\begin{equation}
    C_{ij} =  (\sigma_M^2 + \sigma_\mathrm{SALT}^2 + \sigma_{\mu - z}^2)\delta_{ij}, 
\end{equation}
where $\sigma_M$ is the intrinsic scatter of the \sn\ magnitudes and is also a free parameter of the Hubble diagram fit, $\sigma_\mathrm{SALT}$ is given by Eq.~\ref{eq:salterror}, $\sigma_{\mu - z}$ is given by  Eq.~\ref{eq:rederror}, and $\delta_{ij}$ is the Kronecker-delta.

\subsection{Peculiar velocity contribution to the error budget}\label{Methodsec:PV}
As was pointed out in previous cosmological analyses, in addition to the intrinsic scatter, $\sigma_M$, and standardization, $\sigma_\mathrm{SALT}$, error term, PVs contribute to the error budget \citep{Cooray_Caldwell_2006,davisEffectPeculiarVelocities2011}.
The total covariance for a pair of \sns\ is then
\begin{equation}
    C_{ij} \simeq \left(\frac{5}{c\ln 10}\right)^2\frac{1}{z_i z_j} C_{ij}^{vv}+ (\sigma_M^2 + \sigma_\mathrm{SALT}^2 + \sigma_{\mu - z}^2)\delta_{ij}, 
\end{equation}
where $z_i$ and $z_j$ are the redshifts of the \sns\ and $C_{ij}^{vv}$ is the velocity covariance matrix. 
In this work, we study the impact of different models of the PV covariance matrix, $C_{ij}^{vv}$, on the fit of the Hubble diagram.


A first approach is to completely neglect the contribution of PVs to the error budget; that is, $C_{ij}^{vv}=0$. A second approach is to assume that PVs are uncorrelated and that the velocity covariance is diagonal: 
\begin{equation}
    C_{ij}^{vv} = \sigma_v^2\delta_{ij},
\end{equation}
where $\sigma_v$ is the velocity dispersion.
This error modeling has been used in previous studies: 
\citet{kessler_sdss_2009} used $\sigma_v = 300$ \kms, \cite{betouleImprovedCosmologicalConstraints2014} used $\sigma_v=150$~\kms, and \cite{broutFirstCosmologyResults2019} used $\sigma_v=250$~\kms. In this work, we chose to use this last value of $\sigma_v=250$~\kms.

A more realistic approach is, as was proposed in \cite{davisEffectPeculiarVelocities2011}, to use a complete velocity covariance matrix:  
\begin{equation}
    C_{ij}^{vv} = \frac{H_0^2 f^2}{2\pi^2}\int dk P_{\theta\theta}(k) W(k;\mathbf{x}_i, \mathbf{x}_j),
\end{equation}
where $f$ is the growth rate of structures, $P_{\theta\theta}$ the power spectrum of the velocity divergence, and $W(k;\mathbf{x}_i, \mathbf{x}_j)$ the window function for two \sns\ at positions $\mathbf{x}_i$ and $\mathbf{x}_j$, given by 
\begin{equation}
    \begin{split}
        W_{ij}(k; \mathbf{x}_i, \mathbf{x}_j)         &= \frac{1}{3} \left[ j_0\left(k r_{ij}\right) 
                         - 2j_2\left(k r_{ij}\right) 
                     \right] \cos (\alpha_{ij}) \\
        & \ \ + \frac{x_i x_j }{r_{ij}^2} j_2\left(k r_{ij}\right)\sin^2(\alpha_{ij}),
    \end{split}
\end{equation}
where $r_{ij} = \left \|\mathbf{x}_i - \mathbf{x}_j   \right \|$, $\alpha_{ij}$ is the angle between $\mathbf{x}_i$ and $\mathbf{x}_j$ and $j_n(x)$ are the spherical Bessel functions. 
The covariance matrix was computed using the \texttt{FLIP}\footnote{\url{https://github.com/corentinravoux/flip}} python package. 
In this work, we computed the covariance using the linear velocity-divergence power spectrum, $P_{\theta\theta}^\mathrm{lin}$. In linear theory, due to the continuity equation, $P_{\theta\theta}^\mathrm{lin} = P_{\delta\delta}^\mathrm{lin}$, where $P_{\delta\delta}^\mathrm{lin}$ is the linear density power spectrum. Therefore, we computed the linear density power spectrum using the Boltzmann solver \texttt{CAMB}\footnote{\url{https://camb.info/}} \citep{lewis_camb_2000}. The cosmology parameters used as inputs of the Boltzmann solver are the same as those used to compute $\mu_\mathrm{th}$ (Table.~\ref{tab:siminput} for the simulations and \textit{Planck}18 for the data). Here, we did not attempt to fit for the growth rate, $f$, which is fixed by the fiducial cosmology, and the maximum wave number was chosen as $k_\mathrm{max}=0.2$~\hmpc. 
The diagonal value of the covariance matrix computed as $\sigma_v = \sqrt{C_{ii}^{vv}}$ is $\simeq 260$~\kms\ for the fiducial cosmology used in simulation and $\simeq 280$~\kms\ for the \textit{Planck}18 cosmology. These values are slightly higher but close to the value of 250~\kms\ used in the diagonal velocity error approach.

To evaluate the robustness of our analysis, we varied the assumptions we made to compute the velocity covariance: the input cosmological parameters and the chosen maximum $k$. We found that our analysis is independent of these initial assumptions. A detailed description of this investigation is given in Appendix~\ref{app:systematics}.
 
\section{Test on simulations}\label{simulationresults}

\subsection{Test on a single realization}

We ran a Markov Chain Monte Carlo (MCMC) algorithm using the \texttt{EMCEE}\footnote{\url{ https://emcee.readthedocs.io/}} package to fit the Hubble diagram. Selecting one random realization, we ran three different MCMCs using the different methods to compute the PV covariance matrix. 
The fit was performed maximizing the likelihood in Eq.\ref{eq:HD_likelihood}, where the free parameters are $\alpha$, $\beta$, $M_0$, and $\sigma_m$.

\begin{figure}
    \centering
    \includegraphics[width=\hsize]{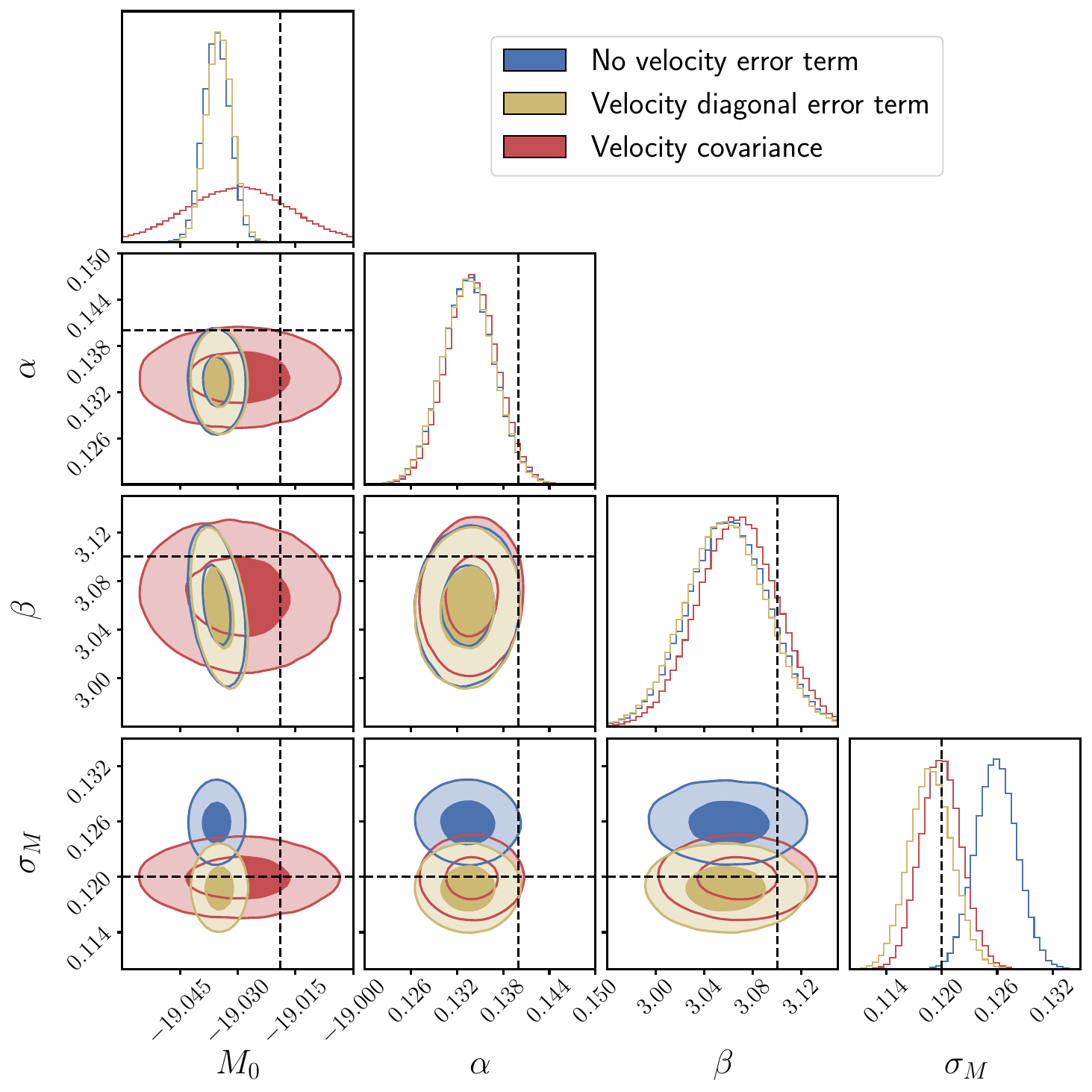}
    \caption{MCMC results of the Hubble diagram fit using the three different methods to compute the PV covariance matrix described in Sect.~\ref{Methodsec}. The fit was performed on one random realization. The dashed black lines show the simulation input value for each parameter. The blue contours show the posterior probability for each parameter when the PV covariance matrix is equal to zero, and PVs are not taken into account. The yellow contours show the results when the covariance matrix is just diagonal, assuming that $\sigma_v = 250$ \kms. Finally, the red contours were obtained when we took into account the PV correlation and the covariance matrix was computed using the linear power spectrum.}
    \label{fig:mcmc_normvsjlavscov}
\end{figure}

Figure~\ref{fig:mcmc_normvsjlavscov} shows the posterior distributions of the fit parameters for a single ZTF survey realization, while Table~\ref{tab:simu1mock} shows the best-fit values of each parameter for the three methods.
The figure shows that we recover the $\alpha$ and $\beta$ parameter input values within $2\sigma$. However, we notice that there is no difference between the three methods in evaluating these two parameters, and that the fitted values are compatible within the error bars, as is shown in Table~\ref{tab:simu1mock}. 

In the case of $M_0$, the PV modeling has a non-negligible impact.  We observe a 4.5~$\sigma$ bias both when PVs are not accounted for or when neglecting their spatial correlations. However, Table~\ref{tab:simu1mock} shows that when accounting for correlations, the uncertainty on $M_0$ is about four times larger than the one obtained in the two other cases, making it compatible with its input value. This suggests that only using a diagonal term might not be sufficient to completely take into account the PVs' impact in a low-redshift sample. We also notice that when using a diagonal term as well as the full covariance we recover the $\sigma_M$ input value, while its value is increased when not taking into account PVs.

When computing the PV contribution to the error budget of $M_0$ as
\begin{equation}
    \sigma_{M_0,\mathrm{pv}} = \sqrt{\sigma_{M_0, \mathrm{vel. \ cov.}}^2 - \sigma_{M_0, \mathrm{no \ vel.}}^2},
    \label{eq:pvcontrib}
\end{equation}
we find that the PVs are the main source of uncertainty for the $M_0$ parameter with $\sigma_{M_0,\mathrm{pv}}\sim 0.013$ mag.

In the next sections, we focus our study on the impact of PVs on the $M_0$ parameter, as the impact on $\alpha$ and $\beta$ is negligible, while $\sigma_M$ is only a nuisance parameter in the fit. As was already mentioned, $M_0$ is directly linked with the intercept of the Hubble diagram, $a_B$, which is the parameter that has to be calibrated through the distance ladder to measure $H_0$. Therefore, the effect of PVs on $M_0$ estimation described in this section could bias the measurement of the Hubble constant when using {\sns}.

\begin{table}[]
\setlength\tabcolsep{4.5pt}
    \centering
    \caption{Results from the MCMC chains obtained on one realization.}
    \begin{tiny}

   \begin{tabular}{c||c|c|c} 
   & No vel. error & Vel. diag. error & Vel. full cov \\[1ex]\hhline{=||=|=|=|}\rule{0pt}{1.5\normalbaselineskip}
        \textbf{$\Delta M_0$} & $-0.0165\pm0.0034$ & $-0.0157\pm0.0033$ & $-0.0108\pm0.0135$\\[1ex]\textbf{$\Delta\alpha$} & $-0.0066\pm0.0034$ & $-0.0066\pm0.0034$ & $-0.0061\pm0.0033$\\[1ex]\textbf{$\Delta\beta$} & $-0.040\pm0.033$ & $-0.043\pm0.034$ & $-0.032\pm0.032$\\[1ex]\textbf{$\Delta\sigma_M$} & $0.0059\pm0.0023$ & $-0.0012\pm0.0023$ & $-0.0001\pm0.0023$\\
        \end{tabular}
    \end{tiny}
    \tablefoot{The values of the parameters, $p$, were computed as the median of the chains and given with respect to the simulation input value, $\Delta p = p_{fit} - p_{true}$. The associated lower and upper error bars were computed as the 16\textsuperscript{th} and 84\textsuperscript{th} percentiles. Since those errors were almost symmetric, we give their means.}    
    \label{tab:simu1mock}
    
\end{table}

\subsection{$M_0$ estimate and uncertainties}\label{sec:sim:M0}
\begin{figure}
    \centering
    \includegraphics[width=\columnwidth]{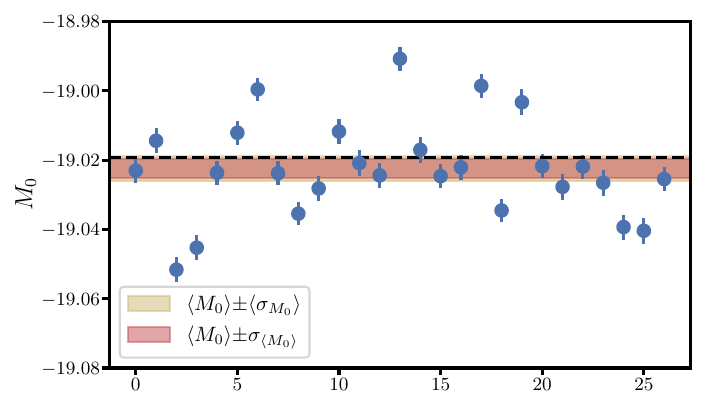}
    \includegraphics[width=\columnwidth]{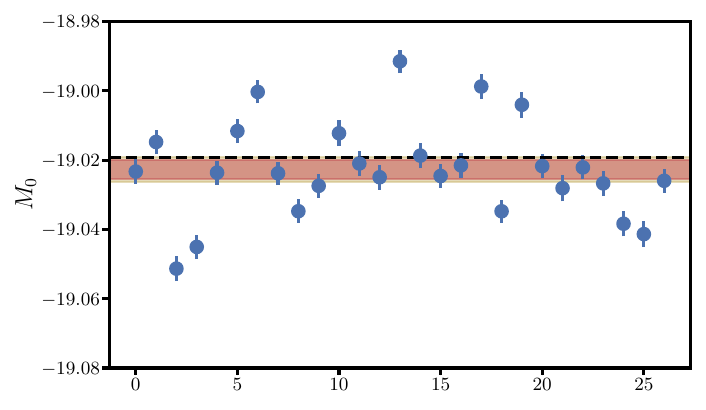}
    \includegraphics[width=\columnwidth]{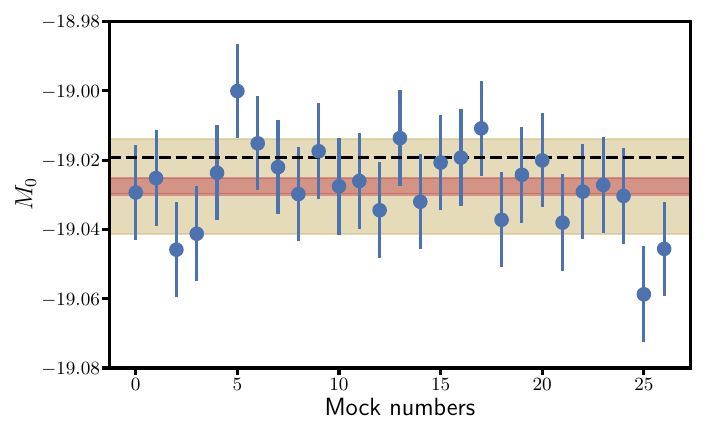}
    \caption{$M_0$ values obtained from the fit of the 27 realizations (realization number 8 is the one used in Fig.~\ref{fig:mcmc_normvsjlavscov}). The blue dots are the results of the fits with error bars from \texttt{iminuit}, while the dashed black line shows the $M_0$ input value of the simulations. The shaded red area is centered on the mean $M_0$ value $\langle M_0 \rangle$ obtained from the 27 realizations and it shows the uncertainty on $\langle M_0 \rangle$. The shaded yellow area is also centered on $\langle M_0 \rangle$, but shows the average uncertainty along the 27 realization results, $\langle\sigma^2_{M_0}\rangle^{1/2}$. \\ \textit{Top panel}: Fit without velocity error term.\\ \textit{Mid panel}: Fit with a velocity diagonal error term.\\ \textit{Bottom panel}: Fit with a full velocity covariance.}
    \label{fig:simM0}
\end{figure}

To see if the parameters values and uncertainties, in particular $M_0$, were well estimated, we ran fits on all our 27 ZTF realizations. We ran the fit using \texttt{iminuit}\footnote{\url{https://iminuit.readthedocs.io/}} \citep{jamesMinuitSystemFunction1975,dembisky_iminuit_2020} to find the maximum of the likelihood. We checked that the results from \texttt{iminuit} are in agreement with the ones from the MCMC and used \texttt{iminuit} as a quicker solution to fit our 27 realizations. Figure~\ref{fig:simM0} shows the results of these fits for $M_0$. We highlight that even for the 27 realization fits, the results for $\alpha$ and $\beta$ do not have any significant variation between the three methods. The additional results for these parameters are presented in Appendix~\ref{app:otherparams}.

\begin{table}
    \centering
    \caption{Comparison between the standard deviation of the parameters results computed across the 27 ZTF realizations and the average uncertainties obtained from the fit.}
    \begin{tiny}
    \begin{tabular}{c||c|c|c|}
                                                           & No vel. error & Vel. diagonal error & Vel. full cov \\[1ex] \hhline{=||=|=|=|}\rule{0pt}{1.5\normalbaselineskip}
         STD($M_0$)                                        & 0.014  & 0.014  & 0.012  \\
         $\langle\sigma_{M_0}^2\rangle^{1/2}$              & 0.004  & 0.004  & 0.014  \\[1ex] \hline\rule{0pt}{1.5\normalbaselineskip}
         STD($\alpha$)                                     & 0.0033 & 0.0032 & 0.0031 \\
         $\langle\sigma_\alpha^2\rangle^{1/2}$             & 0.0035 & 0.0035 & 0.0035 \\[1ex] \hline\rule{0pt}{1.5\normalbaselineskip}
          STD($\beta$)                                     & 0.029  & 0.029  & 0.029  \\
          $\langle\sigma_\beta^2\rangle^{1/2}$             & 0.034  & 0.034  & 0.034  \\[1ex] \hline\rule{0pt}{1.5\normalbaselineskip}
          STD($\sigma_M$)                                  & 0.0034 & 0.0035 & 0.0027 \\
          $\langle\sigma_M\rangle^{1/2}$                   & 0.0023 & 0.0025 & 0.0023 \\
    \end{tabular}
    \end{tiny}
    \label{tab:simscatters}
\end{table}


The top panel of Fig.~\ref{fig:simM0} shows our result for $M_0$ obtained when we performed the fit without using any model for the velocity uncertainties. We measure that the standard deviation between the 27 realizations is $\mathrm{STD}(M_0) = 0.014$ mag, which is three to four times larger than the estimated uncertainty on the average $\langle\sigma_{M_0}^2\rangle^{1/2}= 0.004$ mag. This result shows that uncertainties on $M_0$ are clearly underestimated.
The middle panel of Fig.~\ref{fig:simM0} shows the same results when the contribution of PVs is modeled using the diagonal error term with $\sigma_v = 250$ \kms. The resulting standard deviation over the 27 realizations is $\mathrm{STD}(M_0)=0.014$ mag and the average uncertainty is $\langle\sigma_{M_0}^2\rangle^{1/2}= 0.004$ mag. This result is identical to the previous case. It means that the diagonal error term is not sufficient to model the PV impact.
Finally, the bottom panel of Fig.~\ref{fig:simM0} shows the results for the case in which we use a complete velocity covariance matrix. We obtain a standard deviation of $\mathrm{STD}(M_0)=0.012$ mag with a mean uncertainty of $\langle\sigma_{M_0}^2\rangle^{1/2} = 0.014$ mag. In this last case, the standard deviation and uncertainty are compatible. Applying Eq.~\ref{eq:pvcontrib} used for one realization to the average uncertainties on $M_0$, we retrieve a velocity contribution to the error budget on $M_0$ of $\sigma_{M_0,\mathrm{pv}}\sim 0.013$ mag.

\begin{figure}
    \centering
    \includegraphics[width=\columnwidth]{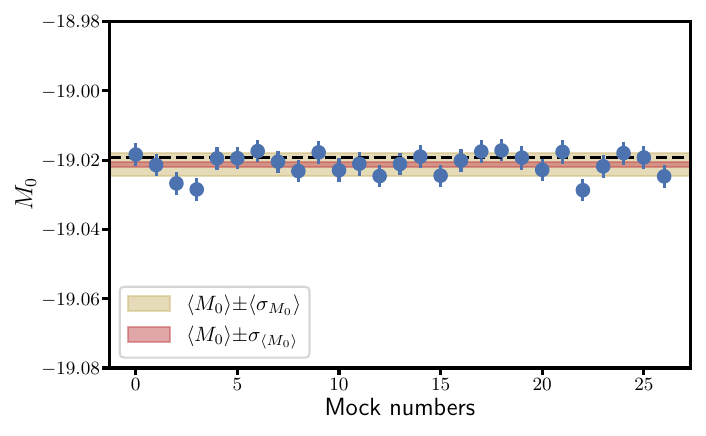}
    \caption{Same as Fig.~\ref{fig:simM0} but the redshifts and magnitudes have been corrected for velocity effects before fitting the Hubble diagram.}
    \label{fig:M0velcorr}
\end{figure}

We notice that in Fig.~\ref{fig:simM0} the central value of $\langle M_0\rangle$ is biased compared to the simulation input value with a difference of $\Delta M_0 = \langle M_0 \rangle - M_{0, \mathrm{fid}} \sim (-8 \pm 2) \times 10^{-3}$ mag. This bias could be the result of correlations of the velocity field between the different mocks that originate from the same N-{\it body} simulated box. In order to study this effect, we ran a fit of the Hubble diagram without the effects of velocities by using the cosmological redshifts, $z_\mathrm{cos}$. The results are shown in Fig.~\ref{fig:M0velcorr}. By correcting the velocities, the bias is reduced to $\Delta M_0 \sim (-2.0 \pm 0.6) \times 10^{-3}$ mag. 
This effect could be explained by the fact that the realizations may be slightly correlated due to large modes inside the full OuterRim box. We understand the remaining small bias to be the first manifestation of the sample selection bias, also known as Malmquist bias, which arises from higher sampling of brighter objects when the distance increases.
However, this bias does not affect the results on one measurement since it is compatible within 1$\sigma$ with respect to the average uncertainty, $\langle \sigma_{M_0}^2\rangle^{1/2}$.

The results of the comparison between the standard deviation of all the fitted parameters across the 27 ZTF survey realizations and the average uncertainties from the fit are summarized in Table~\ref{tab:simscatters}. From these results, we find that in the case of the ZTF sample in the low-redshift regime, diagonal error terms are not sufficient to take into account the impact of PVs. 

\section{The intercept of the Hubble law in ZTF SN~Ia DR2 data}\label{sec:IntercepthubbleDR2}
The increasing uncertainty on $M_0$ that we pointed out in the previous sections could have an impact on Hubble constant measurement from ZTF data \citep{dhawanUniformTypeIa2022}. In such an analysis, $H_0$ is computed using Eq.~\ref{H0-ab}. The two parameters of Eq.~\ref{H0-ab} are the \sns\ absolute magnitude, $M_B$, which has to be calibrated using Cepheids or TRGB, and the intercept of the magnitude-redshift relation, $a_B$. The latter is determined from the \sns\ sample as
\begin{align}
    5a_B &= m_B + \alpha x_1 - \beta c - 5\log D_L(z),
    \label{eq:ab}
\end{align}
were $D_L$ is defined as $D_L = H_{0, \mathrm{fid}} d_{L,\mathrm{fid}}(z)$.
This equation is a reformulation of Eq.~\ref{eq:mu} and the considerations we have made on $M_0$ in the previous section are relevant for $a_B$, since these two parameters are linearly linked as
\begin{equation}
    5a_B = 5\log H_{0,\mathrm{fid}} - M_0 - 25.
\end{equation}
Therefore, the corresponding uncertainty on $H_0$ is 
\begin{equation}
    \frac{\sigma_{H_0}}{H_0} = \frac{\ln10}{5}\sqrt{\sigma_{M_B}^2+\sigma_{5a_B}^2},
    \label{eq:H0err}
\end{equation}
where $\sigma_{M_B}$ is the calibration uncertainty of the absolute magnitude of \sns\ using calibrator stars (e.g., Cepheids or TRGB) and $\sigma_{5a_B}$ is the uncertainty on $5 a_B$.
We ran a MCMC fit using this parameterization of the Hubble diagram on the ZTF SN~Ia DR2 data. 

\begin{figure*}
    \centering
     \includegraphics[width=0.85\textwidth]{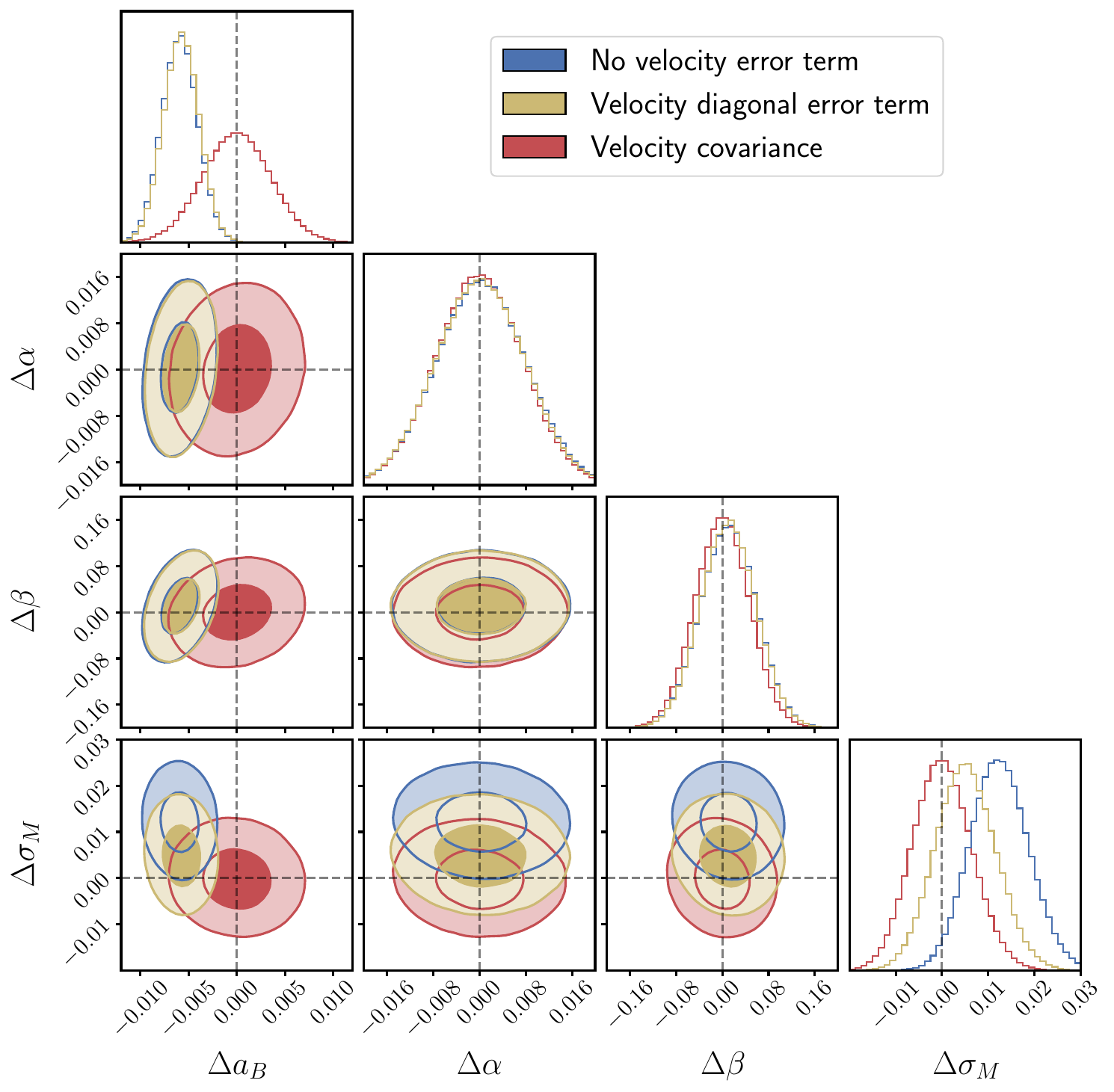}
    \caption{MCMC results of the Hubble diagram fit for the ZTF SN~Ia DR2 data using the three different methods of computing the PV covariance matrix described in Sect. \ref{Methodsec}. The symbols and colors are the same as in Fig.~\ref{fig:mcmc_normvsjlavscov}. The contours are shifted relative to the results obtained using a full covariance matrix for PVs.  }
    \label{fig:mcmc_data}
\end{figure*}

The resulting contours of the MCMC for the three different methods of taking the PV into account are shown in Fig.~\ref{fig:mcmc_data}, and the results are summarized in Table.~\ref{tab:datares}. The values of these results were blinded by the addition of a random offset to the SALT2 magnitude, $m_B$. Therefore, the fitted parameters are shown as differences, $\Delta p$, relative to the results using a full covariance matrix.  As is seen in the simulations, the parameters $\alpha$ and $\beta$ and their corresponding uncertainties are the same for the three methods. The parameter of interest, $a_B$, shifts by about $\sim 0.006$ mag when the PV uncertainties are modeled by the full covariance matrix compared to the case in which PVs are not taken into account or modeled by a diagonal error term. This corresponds to an $H_0$ shift of $\sim 1.0$~\kmsmpc. The uncertainty on $a_B$ doubles, passing from $\sigma_{a_B} = 0.0018$ mag to $\sigma_{a_B} = 0.0035$ mag, when using the full covariance matrix compared to the two other methods.
We also notice that the intrinsic scatter free parameter, $\sigma_M$, increases when using only a diagonal term or when not taking into account velocities because of its degeneracy with the diagonal terms of PV covariance.

We find that, when taking into account PVs with the full covariance matrix, the uncertainty on $a_B$ corresponds to a contribution to the $H_0$ error budget  (Eq.~\ref{eq:H0err}) of $\sigma_{5a_B} \simeq 0.0175$ mag.
Applying Eq.~\ref{eq:pvcontrib} to the uncertainty on $5a_B$, we compute that the velocity contribution is $\sigma_{5a_B, \mathrm{pv}}\sim 0.015$ mag, which is compatible with the result found on $M_0$ with simulations. This could be translated as a contribution of $\sigma_{H_0, \mathrm{pv}} \sim 0.49$ km.s$^{-1}$.Mpc$^{-1}$ on the error on $H_0$.

This result should be compared with the uncertainty on the calibrator term, $\sigma_{M_B}$. In the last measurement of $H_0$ combining TRGB and \sns\ from ZTF \citep{dhawanUniformTypeIa2022}, the uncertainty from calibrators could  be decomposed in three terms that are described in Table~2 of the same paper. The first term is the \sn\ intrinsic scatter, the second term comes from the TRGB magnitude absolute calibration, and the last comes from the scatter of observed TRGB stars in \sn\ host galaxies. In this measurement, the error budget was dominated by 
the \sn\ intrinsic scatter of about $\sim0.15$ mag since there was only one \sn\ in the calibrator sample. In the future, the sample of calibrators is expected to grow up to $\sim100$, decreasing the contribution of \sns\ scatter in calibrator samples to $\sim0.01$ mag. The same is expected for the scattering of TRGB stars in \sns\ host galaxies with an expected uncertainty of $\sim0.005$ mag in future analysis. However, the TRGB absolute calibration uncertainty, which is about $\sim0.038$ mag, is driven by systematic uncertainties. It will be the major contribution to the $H_0$ error budget in future measurements. It could be improved through the use of additional primary anchors and a better control over systematics \cite[see][Sect.~4]{dhawanUniformTypeIa2022} and is expected to be reduced to the level of $\sim0.023$ mag.
To summarize, we could expect an uncertainty on $M_B$ in the range of $\sigma_{M_B} \sim 0.026 - 0.040$ mag. Therefore, considering the expected $\sigma_{M_B}$ and our result for the estimated $\sigma_{5a_B}$ using the DR2 Hubble flow, from Eq.~\ref{eq:H0err} we expect a relative uncertainty on $H_0$ in the range of $1.4\%-2.0\%$ when the velocity covariance matrix is used. This is slightly higher than the relative uncertainty of $1.2\%-1.9\%$ that we computed when not using it. 

\begin{table}[]
    \centering
    \caption{Results from the MCMC chains for the ZTF SN~Ia DR2 data.}
    \begin{tiny}
    \begin{tabular}{c||c||c|c||c|c}
    $p$ & Vel. full cov. & \multicolumn{2}{c||}{No vel. error}      & \multicolumn{2}{c}{Vel. diagonal error}          \\ 
    \cline{2-6}                 & $\sigma_p$   & $\Delta p$ & $\sigma_p$ & $\Delta p$ & $\sigma_p$ \\
    \hhline{=||=||=|=||=|=||}
     $a_B$      & $\pm$ 0.0035 & -0.0059 & $\pm$ 0.0019 & -0.0057 & $\pm$ 0.0018\\ 
     $\alpha$   & $\pm$ 0.0075 & 0.0003  & $\pm$ 0.0078 & 0.0002  & $\pm$ 0.0077\\ 
     $\beta$    & $\pm$ 0.047  & 0.011   & $\pm$ 0.048  & 0.010   & $\pm$ 0.048\\
     $\sigma_M$ & $\pm$ 0.0063 & 0.0124  & $\pm$ 0.0063 & 0.0051  & $\pm$ 0.0065
     \end{tabular}
     \end{tiny}
     \tablefoot{The values and error bars were computed as in Table.\ref{tab:simu1mock}. $\Delta p$ are shifted results with respect to the results obtained with the full covariance matrix and $\sigma_p$ are the error bars obtained for each parameter.}
    \label{tab:datares}
\end{table}

\section{Conclusion}
\label{sec:conclusion}

In this paper, we have used the OuterRim $N$-body simulation described in Sect.~\ref{Datasec} to obtain 27 realisations of the 6-year \sn\ sample of the ZTF survey including PV sample variance.
Most of previous analyses (e.g., \citealt{betouleImprovedCosmologicalConstraints2014} and \citealt{broutFirstCosmologyResults2019}) have used a diagonal error matrix to account for the host-galaxy PVs.  
Due to the low redshift of the ZTF SN~Ia DR2 sample, this approach is not valid to take into account the PV variance. The error bars on the $M_0$ parameter are not compatible with the standard deviation computed across the 27 realisations, either when we did not take into account the PV term or when we used a diagonal matrix to describe PVs, as is shown in Sect.~\ref{simulationresults}. To account for the PV sample variance in the $M_0$ error budget, we need to use a full PV covariance matrix computed using the velocity power spectrum. This result, as in \cite{davisEffectPeculiarVelocities2011}, is an indication that even if in a previous SNe Ia low-z sample the PV correlation was negligible compared to the noise \citep{hutererNoEvidenceBulk2015}, it is no longer valid for a large and homogeneous sample as the ZTF SN~Ia DR2 sample.

We show that these PV correlations also have an impact on the measurement of $H_0$, since they increase the uncertainty on the intercept of the Hubble diagram, $a_B$. We tested this using ZTF SN Ia DR2 data, as is described in Sect.~\ref{sec:IntercepthubbleDR2}. The value of $a_B$ shifts by $\Delta a_B \simeq 0.006$, which corresponds to a shift in the $H_0$ value of about $\simeq 1.0$ \kms$.\mathrm{Mpc}^{-1}$. Moreover, the uncertainty on $a_B$ doubles when using the full PV covariance matrix compared to the other methods. The final contribution of $a_B$ in the $H_0$ error budget is $\sigma_{5a_B} = 0.0175$. The PVs account for $\simeq 0.015$ mag in the value of $\sigma_{5a_B}$ and $\sim 0.49$ km.s$^{-1}$.Mpc$^{-1}$ on the $H_0$ error. This is slightly higher than the result of $\sim 0.31$ km.s$^{-1}$ Mpc$^{-1}$ obtained in \cite{wuSampleVarianceLocal2017}. Therefore, we  conclude that neglecting the non-diagonal terms in the PV covariance matrix brings an underestimation of the uncertainty on the intercept of the Hubble diagram.

The uncertainty on $H_0$ is currently dominated by the systematics from the calibrator sample. Future work on this source of systematics will reduce it, making it even more relevant to correctly take into account the PV contribution to the $H_0$ error budget. Reconstruction methods that correct for PVs are a promising way to account for PV correlations. They should allow one to reduce the scatter due to PVs in the Hubble diagram, and thus suppress the PV sample variance issue. We have shown that an analysis using a full velocity covariance matrix ensures that PV effects are correctly accounted for. While not being competitive, the covariance matrix method can be used as a cross-check to investigate potential biases of the reconstruction methods, where the \sn\ redshifts are corrected using the velocities estimated from the density field \citep{carrickCosmologicalParametersComparison2015,petersonPantheonAnalysisEvaluating2022, carrPantheonAnalysisImproving2022}.

\begin{acknowledgements}
Based on observations obtained with the Samuel Oschin Telescope 48-inch and the 60-inch Telescope at the Palomar Observatory as part of the \textit{Zwicky} Transient Facility project. ZTF is supported by the National Science Foundation under Grants No. AST-1440341 and AST-2034437 and a collaboration including current partners Caltech, IPAC, the Weizmann Institute of Science, the Oskar Klein Center at Stockholm University, the University of Maryland, Deutsches Elektronen-Synchrotron and Humboldt University, the TANGO Consortium of Taiwan, the University of Wisconsin at Milwaukee, Trinity College Dublin, Lawrence Livermore National Laboratories, IN2P3, University of Warwick, Ruhr University Bochum, Northwestern University and former partners the University of Washington, Los Alamos National Laboratories, and Lawrence Berkeley National Laboratories. Operations are conducted by COO, IPAC, and UW.

SED Machine is based upon work supported by the National Science Foundation under Grant No. 1106171 

The project leading to this publication has received funding from Excellence Initiative of Aix-Marseille University - A*MIDEX, a French ``Investissements d'Avenir'' program (AMX-20-CE-02 - DARKUNI). 
This work has been carried out thanks to the support of the DEEPDIP ANR project (ANR-19-CE31-0023).
This work received support from the French government under the France 2030 investment plan, as part of the Initiative d'Excellence d'Aix-Marseille Université - A*MIDEX (AMX-19-IET-008 - IPhU). This project has received funding from the European Research Council (ERC) under the European Union's Horizon 2020 research and innovation programme (grant agreement n°759194 - USNAC).

This project has received funding from the European Research Council (ERC) under the European Union's Horizon 2020 research and innovation programme (grant agreement n°759194 - USNAC)

This work has been supported by the Agence Nationale de la Recherche of the French government through the program ANR-21-CE31-0016-03.

This work has been supported by the research project grant “Understanding the Dynamic Universe” funded by the Knut and Alice Wallenberg Foundation under Dnr KAW 2018.0067,  {\it Vetenskapsr\aa det}, the Swedish Research Council, project 2020-03444 and the G.R.E.A.T research environment, project number 2016-06012.

L. G. acknowledges financial support from the Spanish Ministerio de Ciencia e Innovaci\'on (MCIN) and the Agencia Estatal de Investigaci\'on (AEI) 10.13039/501100011033 under the PID2020-115253GA-I00 HOSTFLOWS project, from Centro Superior de Investigaciones Cient\'ificas (CSIC) under the PIE project 20215AT016 and the program Unidad de Excelencia Mar\'ia de Maeztu CEX2020-001058-M, and from the Departament de Recerca i Universitats de la Generalitat de Catalunya through the 2021-SGR-01270 grant.

G. D. is supported by the H2020 European Research Council grant no. 758638

T. E. M. B. acknowledges financial support from the Spanish Ministerio de Ciencia e Innovaci\'{o}n (MCIN), the Agencia Estatal de Investigaci\'{o}n (AEI) 10.13039/501100011033, and the European Union Next Generation EU/PRTR funds under the 2021 Juan de la Cierva program FJC2021-047124-I and the PID2020-115253GA-I00 HOSTFLOWS project, from Centro Superior de Investigaciones Cient\'{i}ficas (CSIC) under the PIE project 20215AT016, and the program Unidad de Excelencia Mar\'{i}a de Maeztu CEX2020-001058-M.

U. B. and J. H. T. are supported by the H2020 European Research Council grant no. 758638.

\end{acknowledgements}

%
%

\bibliographystyle{aa}
\bibliography{MyBib.bib,newBIB.bib}

\begin{thebibliography}{51}
\expandafter\ifx\csname natexlab\endcsname\relax\def\natexlab#1{#1}\fi

\bibitem[{Barbary {et~al.}(2023)Barbary, Bailey, Barentsen, Barclay, Biswas, Boone, Craig, Feindt, Friesen, Goldstein, Jha, Jones, Mondon, Papadogiannakis, Perrefort, Pierel, Rodney, Rose, Saunders, Sip{\H o}cz, Sofiatti, Thomas, {van Santen}, Vincenzi, Wang, \& {Wood-Vasey}}]{barbarySNCosmo2023}
Barbary, K., Bailey, S., Barentsen, G., {et~al.} 2023, {{SNCosmo}}, Zenodo

\bibitem[{Bellm {et~al.}(2019)Bellm, Kulkarni, Graham, Dekany, Smith, Riddle, Masci, Helou, Prince, Adams, Barbarino, Barlow, Bauer, Beck, Belicki, Biswas, Blagorodnova, Bodewits, Bolin, Brinnel, Brooke, Bue, Bulla, Burruss, Cenko, Chang, Connolly, Coughlin, Cromer, Cunningham, De, Delacroix, Desai, Duev, Eadie, Farnham, Feeney, Feindt, Flynn, Franckowiak, Frederick, Fremling, {Gal-Yam}, Gezari, Giomi, Goldstein, Golkhou, Goobar, Groom, Hacopians, Hale, Henning, Ho, Hover, Howell, Hung, Huppenkothen, Imel, Ip, Ivezi{\'c}, Jackson, Jones, Juric, Kasliwal, Kaspi, Kaye, Kelley, Kowalski, Kramer, Kupfer, Landry, Laher, Lee, Lin, Lin, Lunnan, Giomi, Mahabal, Mao, Miller, Monkewitz, Murphy, Ngeow, Nordin, Nugent, Ofek, Patterson, Penprase, Porter, Rauch, Rebbapragada, Reiley, Rigault, Rodriguez, {van Roestel}, Rusholme, {van Santen}, Schulze, Shupe, Singer, Soumagnac, Stein, Surace, Sollerman, Szkody, Taddia, Terek, Van~Sistine, {van Velzen}, Vestrand, Walters, Ward, Ye, Yu, Yan, \&
  Zolkower}]{bellmZwickyTransientFacility2019}
Bellm, E.~C., Kulkarni, S.~R., Graham, M.~J., {et~al.} 2019, Publications of the Astronomical Society of the Pacific, 131, 018002

\bibitem[{Betoule {et~al.}(2014)Betoule, Kessler, Guy, Mosher, Hardin, Biswas, Astier, {El-Hage}, Konig, Kuhlmann, Marriner, Pain, Regnault, Balland, Bassett, Brown, Campbell, Carlberg, {Cellier-Holzem}, Cinabro, Conley, D'Andrea, DePoy, Doi, Ellis, Fabbro, Filippenko, Foley, Frieman, Fouchez, Galbany, Goobar, Gupta, Hill, Hlozek, Hogan, Hook, Howell, Jha, Guillou, Leloudas, Lidman, Marshall, M{\"o}ller, Mour{\~a}o, Neveu, Nichol, Olmstead, {Palanque-Delabrouille}, Perlmutter, Prieto, Pritchet, Richmond, Riess, {Ruhlmann-Kleider}, Sako, Schahmaneche, Schneider, Smith, Sollerman, Sullivan, Walton, \& Wheeler}]{betouleImprovedCosmologicalConstraints2014}
Betoule, M., Kessler, R., Guy, J., {et~al.} 2014, Astronomy \& Astrophysics, 568, A22

\bibitem[{Blagorodnova {et~al.}(2018)Blagorodnova, Neill, Walters, Kulkarni, Fremling, {Ben-Ami}, Dekany, Fucik, Konidaris, Nash, Ngeow, Ofek, Sullivan, Quimby, Ritter, \& Vyhmeister}]{blagorodnovaSEDMachineRobotic2018}
Blagorodnova, N., Neill, J.~D., Walters, R., {et~al.} 2018, Publications of the Astronomical Society of the Pacific, 130, 035003

\bibitem[{Brout {et~al.}(2019)Brout, Scolnic, Kessler, D'Andrea, Davis, Gupta, Hinton, Kim, Lasker, Lidman, Macaulay, M{\"o}ller, Nichol, Sako, Smith, Sullivan, Zhang, Andersen, Asorey, Avelino, Bassett, Brown, Calcino, Carollo, Challis, Childress, Clocchiatti, Filippenko, Foley, Galbany, Glazebrook, Hoormann, Kasai, Kirshner, Kuehn, Kuhlmann, Lewis, Mandel, March, Miranda, Morganson, Muthukrishna, Nugent, Palmese, Pan, Sharp, Sommer, Swann, Thomas, Tucker, Uddin, Wester, Abbott, Allam, Annis, Avila, Bechtol, Bernstein, Bertin, Brooks, Burke, Rosell, Kind, Carretero, Castander, Cunha, {da Costa}, Davis, De~Vicente, DePoy, Desai, Diehl, Doel, {Drlica-Wagner}, Eifler, Estrada, Fernandez, Flaugher, Fosalba, Frieman, {Garc{\'i}a-Bellido}, Gruen, Gruendl, Gutierrez, Hartley, Hollowood, Honscheid, Hoyle, James, Jarvis, Jeltema, Krause, Lahav, Li, Lima, Maia, Marriner, Marshall, Martini, Menanteau, Miller, Miquel, Ogando, Plazas, Romer, Roodman, Rykoff, Sanchez, Santiago, Scarpine, Schubnell, Serrano,
  {Sevilla-Noarbe}, Smith, {Soares-Santos}, Sobreira, Suchyta, Swanson, Tarle, Thomas, Troxel, Tucker, Vikram, Walker, \& Zhang}]{broutFirstCosmologyResults2019}
Brout, D., Scolnic, D., Kessler, R., {et~al.} 2019, The Astrophysical Journal, 874, 150

\bibitem[{Brout {et~al.}(2022)Brout, Scolnic, Popovic, Riess, Zuntz, Kessler, Carr, Davis, Hinton, Jones, Kenworthy, Peterson, Said, Taylor, Ali, Armstrong, Charvu, Dwomoh, Palmese, Qu, Rose, Stubbs, Vincenzi, Wood, Brown, Chen, Chambers, Coulter, Dai, Dimitriadis, Filippenko, Foley, Jha, Kelsey, Kirshner, M{\"o}ller, Muir, Nadathur, Pan, Rest, {Rojas-Bravo}, Sako, Siebert, Smith, Stahl, \& Wiseman}]{broutPantheonAnalysisCosmological2022}
Brout, D., Scolnic, D., Popovic, B., {et~al.} 2022, The Astrophysical Journal, 938, 110

\bibitem[{Carr {et~al.}(2022)Carr, Davis, Scolnic, Said, Brout, Peterson, \& Kessler}]{carrPantheonAnalysisImproving2022}
Carr, A., Davis, T.~M., Scolnic, D., {et~al.} 2022, Publications of the Astronomical Society of Australia, 39, e046

\bibitem[{Carreres {et~al.}(2023)Carreres, Bautista, Feinstein, Fouchez, Racine, Smith, Amenouche, Aubert, Dhawan, Ginolin, Goobar, Gris, Lacroix, Nuss, Regnault, Rigault, Robert, Rosnet, Sommer, Dekany, Groom, Sravan, Masci, \& Purdum}]{carreresGrowthrateMeasurementTypeIa2023}
Carreres, B., Bautista, J.~E., Feinstein, F., {et~al.} 2023, Astronomy \& Astrophysics, 674, A197

\bibitem[{Carrick {et~al.}(2015)Carrick, Turnbull, Lavaux, \& Hudson}]{carrickCosmologicalParametersComparison2015}
Carrick, J., Turnbull, S.~J., Lavaux, G., \& Hudson, M.~J. 2015, Monthly Notices of the Royal Astronomical Society, 450, 317

\bibitem[{Collaboration {et~al.}(2016)Collaboration, Aghamousa, Aguilar, Ahlen, Alam, Allen, Prieto, Annis, Bailey, Balland, Ballester, Baltay, Beaufore, Bebek, Beers, Bell, Bernal, Besuner, Beutler, Blake, Bleuler, Blomqvist, Blum, Bolton, Briceno, Brooks, Brownstein, Buckley-Geer, Burden, Burtin, Busca, Cahn, Cai, Cardiel-Sas, Carlberg, Carton, Casas, Castander, Cervantes-Cota, Claybaugh, Close, Coker, Cole, Comparat, Cooper, Cousinou, Crocce, Cuby, Cunningham, Davis, Dawson, de~la Macorra, Vicente, Delubac, Derwent, Dey, Dhungana, Ding, Doel, Duan, Ealet, Edelstein, Eftekharzadeh, Eisenstein, Elliott, Escoffier, Evatt, Fagrelius, Fan, Fanning, Farahi, Farihi, Favole, Feng, Fernandez, Findlay, Finkbeiner, Fitzpatrick, Flaugher, Flender, Font-Ribera, Forero-Romero, Fosalba, Frenk, Fumagalli, Gaensicke, Gallo, Garcia-Bellido, Gaztanaga, Fusillo, Gerard, Gershkovich, Giannantonio, Gillet, de~Rivera, Gonzalez-Perez, Gott, Graur, Gutierrez, Guy, Habib, Heetderks, Heetderks, Heitmann, Hellwing, Herrera, Ho,
  Holland, Honscheid, Huff, Hutchinson, Huterer, Hwang, Laguna, Ishikawa, Jacobs, Jeffrey, Jelinsky, Jennings, Jiang, Jimenez, Johnson, Joyce, Jullo, Juneau, Kama, Karcher, Karkar, Kehoe, Kennamer, Kent, Kilbinger, Kim, Kirkby, Kisner, Kitanidis, Kneib, Koposov, Kovacs, Koyama, Kremin, Kron, Kronig, Kueter-Young, Lacey, Lafever, Lahav, Lambert, Lampton, Landriau, Lang, Lauer, Goff, Guillou, Suu, Lee, Lee, Leitner, Lesser, Levi, L'Huillier, Li, Liang, Lin, Linder, Loebman, Lukić, Ma, MacCrann, Magneville, Makarem, Manera, Manser, Marshall, Martini, Massey, Matheson, McCauley, McDonald, McGreer, Meisner, Metcalfe, Miller, Miquel, Moustakas, Myers, Naik, Newman, Nichol, Nicola, da~Costa, Nie, Niz, Norberg, Nord, Norman, Nugent, O'Brien, Oh, Olsen, Padilla, Padmanabhan, Padmanabhan, Palanque-Delabrouille, Palmese, Pappalardo, Pâris, Park, Patej, Peacock, Peiris, Peng, Percival, Perruchot, Pieri, Pogge, Pollack, Poppett, Prada, Prakash, Probst, Rabinowitz, Raichoor, Ree, Refregier, Regal, Reid, Reil, Rezaie,
  Rockosi, Roe, Ronayette, Roodman, Ross, Ross, Rossi, Rozo, Ruhlmann-Kleider, Rykoff, Sabiu, Samushia, Sanchez, Sanchez, Schlegel, Schneider, Schubnell, Secroun, Seljak, Seo, Serrano, Shafieloo, Shan, Sharples, Sholl, Shourt, Silber, Silva, Sirk, Slosar, Smith, Smoot, Som, Song, Sprayberry, Staten, Stefanik, Tarle, Tie, Tinker, Tojeiro, Valdes, Valenzuela, Valluri, Vargas-Magana, Verde, Walker, Wang, Wang, Weaver, Weaverdyck, Wechsler, Weinberg, White, Yang, Yeche, Zhang, Zhao, Zheng, Zhou, Zhou, Zhu, Zou, \& Zu}]{desicollaboration2016desi}
Collaboration, D., Aghamousa, A., Aguilar, J., {et~al.} 2016, The DESI Experiment Part I: Science,Targeting, and Survey Design

\bibitem[{{Cooray} \& {Caldwell}(2006)}]{Cooray_Caldwell_2006}
{Cooray}, A. \& {Caldwell}, R.~R. 2006, \prd, 73, 103002

\bibitem[{Davis {et~al.}(2011)Davis, Hui, Frieman, Haugb{\o}lle, Kessler, Sinclair, Sollerman, Bassett, Marriner, M{\"o}rtsell, Nichol, Richmond, Sako, Schneider, \& Smith}]{davisEffectPeculiarVelocities2011}
Davis, T.~M., Hui, L., Frieman, J.~A., {et~al.} 2011, The Astrophysical Journal, 741, 67

\bibitem[{Dekany {et~al.}(2020)Dekany, Smith, Riddle, Feeney, Porter, Hale, Zolkower, Belicki, Kaye, Henning, Walters, Cromer, Delacroix, Rodriguez, Reiley, Mao, Hover, Murphy, Burruss, Baker, Kowalski, Reif, Mueller, Bellm, Graham, \& Kulkarni}]{dekanyZwickyTransientFacility2020}
Dekany, R., Smith, R.~M., Riddle, R., {et~al.} 2020, Publications of the Astronomical Society of the Pacific, 132, 038001

\bibitem[{Dembinski \& et~al.(2020)}]{dembisky_iminuit_2020}
Dembinski, H. \& et~al., P.~O. 2020

\bibitem[{Dhawan {et~al.}(2022)Dhawan, Goobar, Johansson, Jang, Rigault, Harvey, Maguire, Freedman, Madore, Smith, Sollerman, Kim, Andreoni, Bellm, Coughlin, Dekany, Graham, Kulkarni, Laher, Medford, Neill, Nir, Riddle, \& Rusholme}]{dhawanUniformTypeIa2022}
Dhawan, S., Goobar, A., Johansson, J., {et~al.} 2022, The Astrophysical Journal, 934, 185

\bibitem[{{Graham} {et~al.}(2019){Graham}, {Kulkarni}, {Bellm}, {Adams}, {Barbarino}, {Blagorodnova}, {Bodewits}, {Bolin}, {Brady}, {Cenko}, {Chang}, {Coughlin}, {De}, {Eadie}, {Farnham}, {Feindt}, {Franckowiak}, {Fremling}, {Gezari}, {Ghosh}, {Goldstein}, {Golkhou}, {Goobar}, {Ho}, {Huppenkothen}, {Ivezi{\'c}}, {Jones}, {Juric}, {Kaplan}, {Kasliwal}, {Kelley}, {Kupfer}, {Lee}, {Lin}, {Lunnan}, {Mahabal}, {Miller}, {Ngeow}, {Nugent}, {Ofek}, {Prince}, {Rauch}, {van Roestel}, {Schulze}, {Singer}, {Sollerman}, {Taddia}, {Yan}, {Ye}, {Yu}, {Barlow}, {Bauer}, {Beck}, {Belicki}, {Biswas}, {Brinnel}, {Brooke}, {Bue}, {Bulla}, {Burruss}, {Connolly}, {Cromer}, {Cunningham}, {Dekany}, {Delacroix}, {Desai}, {Duev}, {Feeney}, {Flynn}, {Frederick}, {Gal-Yam}, {Giomi}, {Groom}, {Hacopians}, {Hale}, {Helou}, {Henning}, {Hover}, {Hillenbrand}, {Howell}, {Hung}, {Imel}, {Ip}, {Jackson}, {Kaspi}, {Kaye}, {Kowalski}, {Kramer}, {Kuhn}, {Landry}, {Laher}, {Mao}, {Masci}, {Monkewitz}, {Murphy}, {Nordin}, {Patterson}, {Penprase},
  {Porter}, {Rebbapragada}, {Reiley}, {Riddle}, {Rigault}, {Rodriguez}, {Rusholme}, {van Santen}, {Shupe}, {Smith}, {Soumagnac}, {Stein}, {Surace}, {Szkody}, {Terek}, {Van Sistine}, {van Velzen}, {Vestrand}, {Walters}, {Ward}, {Zhang}, \& {Zolkower}}]{graham_ztf_2019}
{Graham}, M.~J., {Kulkarni}, S.~R., {Bellm}, E.~C., {et~al.} 2019, \pasp, 131, 078001

\bibitem[{Graziani {et~al.}(2019)Graziani, Courtois, Lavaux, Hoffman, Tully, Copin, \& Pomar{\`e}de}]{grazianiPeculiarVelocityField2019}
Graziani, R., Courtois, H.~M., Lavaux, G., {et~al.} 2019, Monthly Notices of the Royal Astronomical Society, 488, 5438

\bibitem[{Guy {et~al.}(2007)Guy, Astier, Baumont, Hardin, Pain, Regnault, Basa, Carlberg, Conley, Fabbro, Fouchez, Hook, Howell, Perrett, Pritchet, Rich, Sullivan, Antilogus, Aubourg, Bazin, Bronder, Filiol, {Palanque-Delabrouille}, Ripoche, \& {Ruhlmann-Kleider}}]{guySALT2UsingDistant2007}
Guy, J., Astier, P., Baumont, S., {et~al.} 2007, Astronomy \& Astrophysics, 466, 11

\bibitem[{Guy {et~al.}(2010)Guy, Sullivan, Conley, Regnault, Astier, Balland, Basa, Carlberg, Fouchez, Hardin, Hook, Howell, Pain, {Palanque-Delabrouille}, Perrett, Pritchet, Rich, {Ruhlmann-Kleider}, Balam, Baumont, Ellis, Fabbro, Fakhouri, Fourmanoit, {Gonzalez-Gaitan}, Graham, Hsiao, Kronborg, Lidman, Mourao, Perlmutter, Ripoche, Suzuki, \& Walker}]{guySupernovaLegacySurvey2010}
Guy, J., Sullivan, M., Conley, A., {et~al.} 2010, Astronomy \& Astrophysics, 523, A7

\bibitem[{Heitmann {et~al.}(2019)Heitmann, Finkel, Pope, Morozov, Frontiere, Habib, Rangel, Uram, Korytov, Child, Flender, Insley, \& Rizzi}]{heitmannOuterRimSimulation2019}
Heitmann, K., Finkel, H., Pope, A., {et~al.} 2019, The Astrophysical Journal Supplement Series, 245, 16

\bibitem[{Howlett {et~al.}(2017)Howlett, {Staveley-Smith}, Elahi, Hong, Jarrett, Jones, Koribalski, Macri, Masters, \& Springob}]{howlett2MTFVIMeasuring2017}
Howlett, C., {Staveley-Smith}, L., Elahi, P.~J., {et~al.} 2017, Monthly Notices of the Royal Astronomical Society, 471, 3135

\bibitem[{Hui \& Greene(2006)}]{huiCorrelatedFluctuationsLuminosity2006}
Hui, L. \& Greene, P.~B. 2006, Physical Review D, 73, 123526

\bibitem[{Huterer {et~al.}(2015)Huterer, Shafer, \& Schmidt}]{hutererNoEvidenceBulk2015}
Huterer, D., Shafer, D.~L., \& Schmidt, F. 2015, Journal of Cosmology and Astroparticle Physics, 2015, 033

\bibitem[{James \& Roos(1975)}]{jamesMinuitSystemFunction1975}
James, F. \& Roos, M. 1975, Computer Physics Communications, 10, 343

\bibitem[{Johnson {et~al.}(2014)Johnson, Blake, Koda, Ma, Colless, Crocce, Davis, Jones, Lucey, Magoulas, Mould, Scrimgeour, \& Springob}]{johnson6dFGalaxyVelocity2014}
Johnson, A., Blake, C., Koda, J., {et~al.} 2014, Monthly Notices of the Royal Astronomical Society, 444, 3926

\bibitem[{Kenworthy {et~al.}(2022)Kenworthy, Riess, Scolnic, Yuan, Bernal, Brout, Casertano, Jones, Macri, \& Peterson}]{Darcy_h02rungs_2022}
Kenworthy, W.~D., Riess, A.~G., Scolnic, D., {et~al.} 2022, The Astrophysical Journal, 935, 83

\bibitem[{{Kessler} {et~al.}(2009){Kessler}, {Becker}, {Cinabro}, {Vanderplas}, {Frieman}, {Marriner}, {Davis}, {Dilday}, {Holtzman}, {Jha}, {Lampeitl}, {Sako}, {Smith}, {Zheng}, {Nichol}, {Bassett}, {Bender}, {Depoy}, {Doi}, {Elson}, {Filippenko}, {Foley}, {Garnavich}, {Hopp}, {Ihara}, {Ketzeback}, {Kollatschny}, {Konishi}, {Marshall}, {McMillan}, {Miknaitis}, {Morokuma}, {M{\"o}rtsell}, {Pan}, {Prieto}, {Richmond}, {Riess}, {Romani}, {Schneider}, {Sollerman}, {Takanashi}, {Tokita}, {van der Heyden}, {Wheeler}, {Yasuda}, \& {York}}]{kessler_sdss_2009}
{Kessler}, R., {Becker}, A.~C., {Cinabro}, D., {et~al.} 2009, \apjs, 185, 32

\bibitem[{{Kim} {et~al.}(2022){Kim}, {Rigault}, {Neill}, {Briday}, {Copin}, {Lezmy}, {Nicolas}, {Riddle}, {Sharma}, {Smith}, {Sollerman}, \& {Walters}}]{KimNewModSEDm}
{Kim}, Y.~L., {Rigault}, M., {Neill}, J.~D., {et~al.} 2022, \pasp, 134, 024505

\bibitem[{Lai {et~al.}(2023)Lai, Howlett, \& Davis}]{laiUsingPeculiarVelocity2023}
Lai, Y., Howlett, C., \& Davis, T.~M. 2023, Monthly Notices of the Royal Astronomical Society, 518, 1840

\bibitem[{Laureijs {et~al.}(2011)Laureijs, Amiaux, Arduini, Auguères, Brinchmann, Cole, Cropper, Dabin, Duvet, Ealet, Garilli, Gondoin, Guzzo, Hoar, Hoekstra, Holmes, Kitching, Maciaszek, Mellier, Pasian, Percival, Rhodes, Criado, Sauvage, Scaramella, Valenziano, Warren, Bender, Castander, Cimatti, Fèvre, Kurki-Suonio, Levi, Lilje, Meylan, Nichol, Pedersen, Popa, Lopez, Rix, Rottgering, Zeilinger, Grupp, Hudelot, Massey, Meneghetti, Miller, Paltani, Paulin-Henriksson, Pires, Saxton, Schrabback, Seidel, Walsh, Aghanim, Amendola, Bartlett, Baccigalupi, Beaulieu, Benabed, Cuby, Elbaz, Fosalba, Gavazzi, Helmi, Hook, Irwin, Kneib, Kunz, Mannucci, Moscardini, Tao, Teyssier, Weller, Zamorani, Osorio, Boulade, Foumond, Giorgio, Guttridge, James, Kemp, Martignac, Spencer, Walton, Blümchen, Bonoli, Bortoletto, Cerna, Corcione, Fabron, Jahnke, Ligori, Madrid, Martin, Morgante, Pamplona, Prieto, Riva, Toledo, Trifoglio, Zerbi, Abdalla, Douspis, Grenet, Borgani, Bouwens, Courbin, Delouis, Dubath, Fontana, Frailis,
  Grazian, Koppenhöfer, Mansutti, Melchior, Mignoli, Mohr, Neissner, Noddle, Poncet, Scodeggio, Serrano, Shane, Starck, Surace, Taylor, Verdoes-Kleijn, Vuerli, Williams, Zacchei, Altieri, Sanz, Kohley, Oosterbroek, Astier, Bacon, Bardelli, Baugh, Bellagamba, Benoist, Bianchi, Biviano, Branchini, Carbone, Cardone, Clements, Colombi, Conselice, Cresci, Deacon, Dunlop, Fedeli, Fontanot, Franzetti, Giocoli, Garcia-Bellido, Gow, Heavens, Hewett, Heymans, Holland, Huang, Ilbert, Joachimi, Jennins, Kerins, Kiessling, Kirk, Kotak, Krause, Lahav, van Leeuwen, Lesgourgues, Lombardi, Magliocchetti, Maguire, Majerotto, Maoli, Marulli, Maurogordato, McCracken, McLure, Melchiorri, Merson, Moresco, Nonino, Norberg, Peacock, Pello, Penny, Pettorino, Porto, Pozzetti, Quercellini, Radovich, Rassat, Roche, Ronayette, Rossetti, Sartoris, Schneider, Semboloni, Serjeant, Simpson, Skordis, Smadja, Smartt, Spano, Spiro, Sullivan, Tilquin, Trotta, Verde, Wang, Williger, Zhao, Zoubian, \& Zucca}]{laureijs2011euclid}
Laureijs, R., Amiaux, J., Arduini, S., {et~al.} 2011, Euclid Definition Study Report

\bibitem[{Lavaux \& Hudson(2011)}]{lavaux2MGalaxyRedshift2011}
Lavaux, G. \& Hudson, M.~J. 2011, Monthly Notices of the Royal Astronomical Society, 416, 2840

\bibitem[{{Lewis} {et~al.}(2000){Lewis}, {Challinor}, \& {Lasenby}}]{lewis_camb_2000}
{Lewis}, A., {Challinor}, A., \& {Lasenby}, A. 2000, \apj, 538, 473

\bibitem[{Lilow \& Nusser(2021)}]{Lilow2021}
Lilow, R. \& Nusser, A. 2021, Monthly Notices of the Royal Astronomical Society, 507, 1557

\bibitem[{{LSST Science Collaboration} {et~al.}(2009){LSST Science Collaboration}, {Abell}, {Allison}, {Anderson}, {Andrew}, {Angel}, {Armus}, {Arnett}, {Asztalos}, {Axelrod}, {Bailey}, {Ballantyne}, {Bankert}, {Barkhouse}, {Barr}, {Barrientos}, {Barth}, {Bartlett}, {Becker}, {Becla}, {Beers}, {Bernstein}, {Biswas}, {Blanton}, {Bloom}, {Bochanski}, {Boeshaar}, {Borne}, {Bradac}, {Brandt}, {Bridge}, {Brown}, {Brunner}, {Bullock}, {Burgasser}, {Burge}, {Burke}, {Cargile}, {Chandrasekharan}, {Chartas}, {Chesley}, {Chu}, {Cinabro}, {Claire}, {Claver}, {Clowe}, {Connolly}, {Cook}, {Cooke}, {Cooray}, {Covey}, {Culliton}, {de Jong}, {de Vries}, {Debattista}, {Delgado}, {Dell'Antonio}, {Dhital}, {Di Stefano}, {Dickinson}, {Dilday}, {Djorgovski}, {Dobler}, {Donalek}, {Dubois-Felsmann}, {Durech}, {Eliasdottir}, {Eracleous}, {Eyer}, {Falco}, {Fan}, {Fassnacht}, {Ferguson}, {Fernandez}, {Fields}, {Finkbeiner}, {Figueroa}, {Fox}, {Francke}, {Frank}, {Frieman}, {Fromenteau}, {Furqan}, {Galaz}, {Gal-Yam}, {Garnavich},
  {Gawiser}, {Geary}, {Gee}, {Gibson}, {Gilmore}, {Grace}, {Green}, {Gressler}, {Grillmair}, {Habib}, {Haggerty}, {Hamuy}, {Harris}, {Hawley}, {Heavens}, {Hebb}, {Henry}, {Hileman}, {Hilton}, {Hoadley}, {Holberg}, {Holman}, {Howell}, {Infante}, {Ivezic}, {Jacoby}, {Jain}, {R}, {Jedicke}, {Jee}, {Garrett Jernigan}, {Jha}, {Johnston}, {Jones}, {Juric}, {Kaasalainen}, {Styliani}, {Kafka}, {Kahn}, {Kaib}, {Kalirai}, {Kantor}, {Kasliwal}, {Keeton}, {Kessler}, {Knezevic}, {Kowalski}, {Krabbendam}, {Krughoff}, {Kulkarni}, {Kuhlman}, {Lacy}, {Lepine}, {Liang}, {Lien}, {Lira}, {Long}, {Lorenz}, {Lotz}, {Lupton}, {Lutz}, {Macri}, {Mahabal}, {Mandelbaum}, {Marshall}, {May}, {McGehee}, {Meadows}, {Meert}, {Milani}, {Miller}, {Miller}, {Mills}, {Minniti}, {Monet}, {Mukadam}, {Nakar}, {Neill}, {Newman}, {Nikolaev}, {Nordby}, {O'Connor}, {Oguri}, {Oliver}, {Olivier}, {Olsen}, {Olsen}, {Olszewski}, {Oluseyi}, {Padilla}, {Parker}, {Pepper}, {Peterson}, {Petry}, {Pinto}, {Pizagno}, {Popescu}, {Prsa}, {Radcka}, {Raddick},
  {Rasmussen}, {Rau}, {Rho}, {Rhoads}, {Richards}, {Ridgway}, {Robertson}, {Roskar}, {Saha}, {Sarajedini}, {Scannapieco}, {Schalk}, {Schindler}, {Schmidt}, {Schmidt}, {Schneider}, {Schumacher}, {Scranton}, {Sebag}, {Seppala}, {Shemmer}, {Simon}, {Sivertz}, {Smith}, {Allyn Smith}, {Smith}, {Spitz}, {Stanford}, {Stassun}, {Strader}, {Strauss}, {Stubbs}, {Sweeney}, {Szalay}, {Szkody}, {Takada}, {Thorman}, {Trilling}, {Trimble}, {Tyson}, {Van Berg}, {Vanden Berk}, {VanderPlas}, {Verde}, {Vrsnak}, {Walkowicz}, {Wandelt}, {Wang}, {Wang}, {Warner}, {Wechsler}, {West}, {Wiecha}, {Williams}, {Willman}, {Wittman}, {Wolff}, {Wood-Vasey}, {Wozniak}, {Young}, {Zentner}, \& {Zhan}}]{LSST_2009_sciencebook}
{LSST Science Collaboration}, {Abell}, P.~A., {Allison}, J., {et~al.} 2009, arXiv e-prints, arXiv:0912.0201

\bibitem[{{Masci} {et~al.}(2019){Masci}, {Laher}, {Rusholme}, {Shupe}, {Groom}, {Surace}, {Jackson}, {Monkewitz}, {Beck}, {Flynn}, {Terek}, {Landry}, {Hacopians}, {Desai}, {Howell}, {Brooke}, {Imel}, {Wachter}, {Ye}, {Lin}, {Cenko}, {Cunningham}, {Rebbapragada}, {Bue}, {Miller}, {Mahabal}, {Bellm}, {Patterson}, {Juri{\'c}}, {Golkhou}, {Ofek}, {Walters}, {Graham}, {Kasliwal}, {Dekany}, {Kupfer}, {Burdge}, {Cannella}, {Barlow}, {Van Sistine}, {Giomi}, {Fremling}, {Blagorodnova}, {Levitan}, {Riddle}, {Smith}, {Helou}, {Prince}, \& {Kulkarni}}]{masci_2019_ztf}
{Masci}, F.~J., {Laher}, R.~R., {Rusholme}, B., {et~al.} 2019, \pasp, 131, 018003

\bibitem[{Nicolas {et~al.}(2021)Nicolas, Rigault, Copin, Graziani, Aldering, Briday, Nordin, Kim, Perlmutter, \& Smith}]{nicolasRedshiftEvolutionUnderlying2021}
Nicolas, N., Rigault, M., Copin, Y., {et~al.} 2021, Astronomy \& Astrophysics, 649, A74

\bibitem[{{Peebles}(1980)}]{peebles_LSS}
{Peebles}, P.~J.~E. 1980, {The large-scale structure of the universe}

\bibitem[{Perley {et~al.}(2020)Perley, Fremling, Sollerman, Miller, Dahiwale, Sharma, Bellm, Biswas, Brink, Bruch, De, Dekany, Drake, Duev, Filippenko, {Gal-Yam}, Goobar, Graham, Graham, Ho, Irani, Kasliwal, Kim, Kulkarni, Mahabal, Masci, Modak, Neill, Nordin, Riddle, Soumagnac, Strotjohann, Schulze, Taggart, Tzanidakis, Walters, \& Yan}]{perleyZwickyTransientFacility2020}
Perley, D.~A., Fremling, C., Sollerman, J., {et~al.} 2020, The Astrophysical Journal, 904, 35

\bibitem[{Peterson {et~al.}(2022)Peterson, Kenworthy, Scolnic, Riess, Brout, Carr, Courtois, Davis, Dwomoh, Jones, Popovic, Rose, \& Said}]{petersonPantheonAnalysisEvaluating2022}
Peterson, E.~R., Kenworthy, W.~D., Scolnic, D., {et~al.} 2022, The Astrophysical Journal, 938, 112

\bibitem[{{Planck Collaboration} {et~al.}(2020){Planck Collaboration}, Elsner, En{\ss}lin, Eriksen, Fantaye, Farhang, Fergusson, {Fernandez-Cobos}, Finelli, Forastieri, Frailis, Franceschi, Frolov, Galeotta, Galli, Ganga, {G{\'e}nova-Santos}, Gerbino, Ghosh, {Gonz{\'a}lez-Nuevo}, G{\'o}rski, Gratton, Gruppuso, Gudmundsson, Hamann, Handley, Hansen, Herranz, Hildebrandt, Hivon, Jaffe, Jones, Karakci, Keih{\"a}nen, Keskitalo, Kiiveri, Kim, Kisner, Knox, Krachmalnicoff, Kunz, {Kurki-Suonio}, Lagache, Lamarre, Lasenby, Lattanzi, Lawrence, Jeune, Lemos, Lesgourgues, Levrier, Lewis, Liguori, Lilje, Lilley, Lindholm, {L{\'o}pez-Caniego}, Lubin, Ma, {Mac{\'i}as-P{\'e}rez}, Maggio, Maino, Mandolesi, Mangilli, {Marcos-Caballero}, Maris, Martin, Martinelli, {Mart{\'i}nez-Gonz{\'a}lez}, Matarrese, Mauri, McEwen, Meinhold, Melchiorri, Migliaccio, Millea, Mitra, {Miville-Desch{\^e}nes}, Montier, Morgante, Natoli, {N{\o}rgaard-Nielsen}, Pagano, Paoletti, Patanchon, Perrotta, Pettorino, Piacentini, Polastri, Polenta, Puget,
  Rachen, Reinecke, Remazeilles, Renzi, Rocha, Rosset, Roudier, {Rubi{\~n}o-Mart{\'i}n}, {Ruiz-Granados}, Salvati, Sandri, Savelainen, Scott, Shellard, Sirignano, Sirri, Spencer, Sunyaev, {Suur-Uski}, Tauber, Tavagnacco, Tenti, Toffolatti, Tomasi, Trombetti, Valenziano, Valiviita, Van~Tent, Vibert, Vielva, Villa, Vittorio, Wandelt, Wehus, White, \& White}]{planckcollaborationPlanck2018Results2020}
{Planck Collaboration}, Elsner, F., En{\ss}lin, T.~A., {et~al.} 2020, Astronomy \& Astrophysics, 641, A6

\bibitem[{Riess {et~al.}(2022)Riess, Yuan, Macri, Scolnic, Brout, Casertano, Jones, Murakami, Breuval, Brink, Filippenko, Hoffmann, Jha, Kenworthy, Anand, Mackenty, Stahl, \& Zheng}]{riessComprehensiveMeasurementLocal2022}
Riess, A.~G., Yuan, W., Macri, L.~M., {et~al.} 2022, The Astrophysical Journal Letters, 934, L7

\bibitem[{Rigault {et~al.}(2024)Rigault, Smith, \& SWG}]{rigaultZTFoverview}
Rigault, Smith, \& SWG, Z. 2024, submitted (not yet accepted) to Astronomy \& Astrophysics

\bibitem[{Rigault {et~al.}(2019)Rigault, Neill, Blagorodnova, Dugas, Feeney, Walters, Brinnel, Copin, Fremling, Nordin, \& Sollerman}]{rigaultFullyAutomatedIntegral2019}
Rigault, M., Neill, J.~D., Blagorodnova, N., {et~al.} 2019, Astronomy and Astrophysics, 627, A115

\bibitem[{Said {et~al.}(2020)Said, Colless, Magoulas, Lucey, \& Hudson}]{saidJointAnalysis6dFGS2020}
Said, K., Colless, M., Magoulas, C., Lucey, J.~R., \& Hudson, M.~J. 2020, Monthly Notices of the Royal Astronomical Society, 497, 1275

\bibitem[{Scolnic \& Kessler(2016)}]{scolnicMeasuringTypeIa2016}
Scolnic, D. \& Kessler, R. 2016, The Astrophysical Journal, 822, L35

\bibitem[{Scolnic {et~al.}(2023)Scolnic, Riess, Wu, Li, Anand, Beaton, Casertano, Anderson, Dhawan, \& Ke}]{scolnic2023cats}
Scolnic, D., Riess, A.~G., Wu, J., {et~al.} 2023, CATS: The Hubble Constant from Standardized TRGB and Type Ia Supernova Measurements

\bibitem[{Smith {et~al.}(2024)Smith, Rigault, \& SWG}]{smith}
Smith, Rigault, \& SWG, Z. 2024

\bibitem[{Soumagnac {et~al.}(2024)Soumagnac, Nugent, Knop, Ho, Hohensee, Awbrey, Andersen, Aldering, Ventura, Aguilar, Ahlen, Benzvi, Brooks, Brout, Claybaugh, Davis, Dawson, de~la Macorra, Dey, Dey, Doel, Douglass, Forero-Romero, Gaztanaga, Gontcho, Graur, Guy, Hahn, Honscheid, Howlett, Kim, Kisner, Kremin, Lambert, Landriau, Lang, Guillou, Manera, Meisner, Miquel, Moustakas, Myers, Nie, Palmese, Parkinson, Poppett, Prada, Qin, Rezaie, Rossi, Sanchez, Schlegel, Schubnell, Silber, Tarle, Weaver, \& Zhou}]{Maayane2024}
Soumagnac, M.~T., Nugent, P., Knop, R.~A., {et~al.} 2024 [\eprint{arXiv:2405.03857}]

\bibitem[{Tripp(1998)}]{trippTwoparameterLuminosityCorrection1998}
Tripp, R. 1998, Astronomy and Astrophysics, 331, 815

\bibitem[{Turner {et~al.}(2022)Turner, Blake, \& Ruggeri}]{turnerLocalMeasurementGrowth2022}
Turner, R.~J., Blake, C., \& Ruggeri, R. 2022, A Local Measurement of the Growth Rate from Peculiar Velocities and Galaxy Clustering Correlations in the {{6dF Galaxy Survey}}

\bibitem[{Wu \& Huterer(2017)}]{wuSampleVarianceLocal2017}
Wu, H.-Y. \& Huterer, D. 2017, Monthly Notices of the Royal Astronomical Society, 471, 4946

\end{thebibliography}

\begin{appendix}
\section{Discussion on correlation between sub-boxes}\label{app:subboxescorr}
\begin{figure}
    \centering
    \includegraphics[width=\columnwidth]{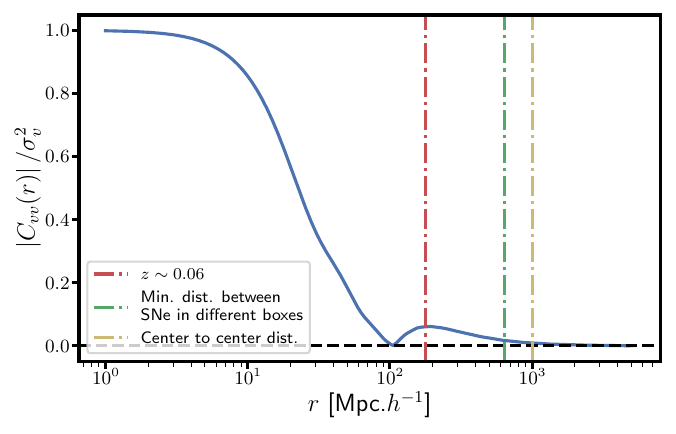}
    \caption{Peculiar velocity correlation absolute value between two galaxies on the same line-of-sight in function of the radial separation. The vertical dotted lines indicate the distance corresponding to the redshift cut $z\sim 0.06$ (red), the minimum distance between two SNe Ia that pass the redshift cut in two different sub-boxes (green) and the distance between two sub-boxes center (yellow).}
    \label{fig:corrfun}
\end{figure}

We produced several realizations of ZTF by cutting a single OuterRim simulation into non-overlapping sub-boxes. However, PVs are coherent over large scales, so these realizations are not fully independent. We estimated the correlation function between PVs inside different sub-boxes as a function of scale, shown in Fig.~\ref{fig:corrfun}. We see that the absolute value of the correlation between two SN Ia host that pass the redshift cut of $z < 0.06$, that corresponds to a distance of $\sim 180$ Mpc.$h^{-1}$ in our fiducial cosmology, in two different sub-boxes is $\sim 0.016$. The correlation between velocities at a distance from one sub-box center to another sub-center is $\sim 0.008$. From the comparison of these values to the average correlation absolute value between two SNe Ia that pass the redshift cut inside one mock $\sim 0.14$, we conclude that the correlation between the sub-boxes can be mostly neglected. However, we see in Sect.~\ref{sec:sim:M0} that a part of the small bias observed on $M_0$ could be explained by the effect of the velocity correlation across the different sub-boxes since removing velocities seems to lower the bias.

\section{Testing the assumptions in the covariance computation}\label{app:systematics}

We assert the robustness of the analysis by testing it against the assumptions we have done to compute the velocity covariance. In particular, the input cosmological parameters in Appendix~\ref{app:changecosmology} and the chosen maximum scale in Appendix~\ref{app:changekmax}. During the analysis we have assumed a linear power spectrum $P_{\theta\theta}^\mathrm{lin}$ in order to compute the complete velocity covariance. This choice is done because in this work we are not measuring the parameters of the cosmological model. Here, we are analyzing systematics and the choice of the power spectrum is a second order effect. Therefore, for simplicity we choose to utilize the linear power spectrum. We highlight that we perform few test in order to assess the impact of the power spectrum choice. The results, which are not shown in the paper, are that changing the power spectrum does not change at all the final results presented in this work. For an extensive description on the non linear power spectrum models and the differences with the linear one see \citet{carreresGrowthrateMeasurementTypeIa2023} and references therein.

\subsection{Changing the input cosmology}\label{app:changecosmology}

\begin{figure}
    \centering
    \includegraphics[width=\hsize]{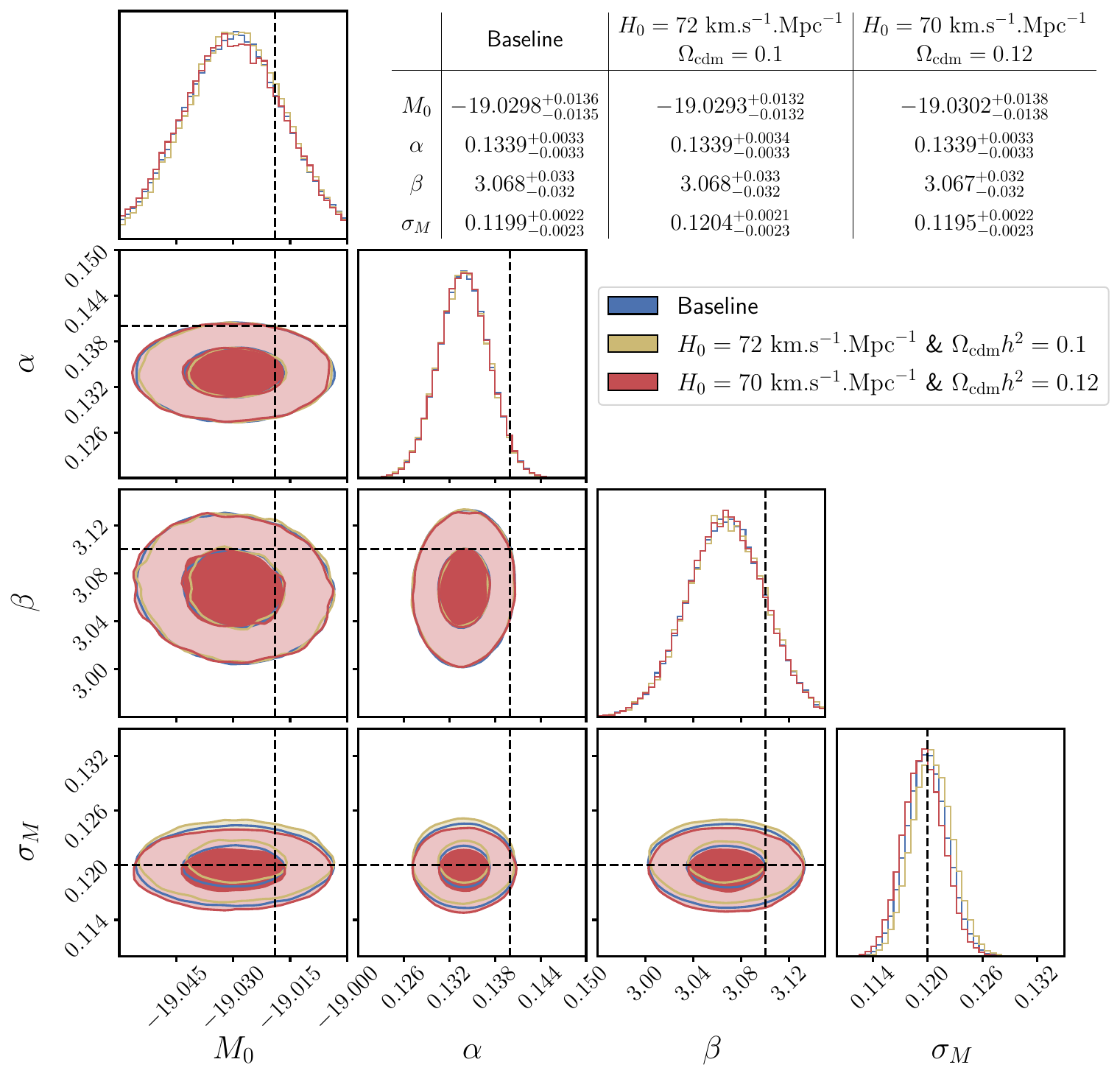}
    \caption{MCMC results of the Hubble diagram fit using three different values input cosmologies to compute the power spectrum $P_{\theta\theta}$ for the complete velocity covariance. The fit is performed on the same realization shown in Fig.~\ref{fig:mcmc_normvsjlavscov}. The dashed black lines show the simulation input value for each parameter. The baseline refers to OuterRim cosmology as described in Sect. \ref{Datasec}. The table shows the median posterior probability, the fitted value, for each parameter and the relative error.}
    \label{fig:mcmc_diffcosmo_cov}
\end{figure}

As a first test we perform the Hubble diagram fit on one random realization changing the input cosmology in the covariance computation respect to the one presented in Sect. \ref{Datasec}. The results of the MCMC chains are shown in Fig.~\ref{fig:mcmc_diffcosmo_cov}. The figure shows that changing $H_0$ and the density of cold dark matter, $\Omega_c$, does not change the results of the Hubble diagram fit. Furthermore, Fig.~\ref{fig:mcmc_diffcosmo_cov} table shows that the recovered parameter values for the three different cosmology are exactly the same within the uncertainties. Therefore, we conclude that the results presented in this work are independent from the choice of the cosmological parameters in the velocity covariance computation. 

\subsection{Changing the maximum scale}\label{app:changekmax}

\begin{figure}
    \centering
    \includegraphics[width=\hsize]{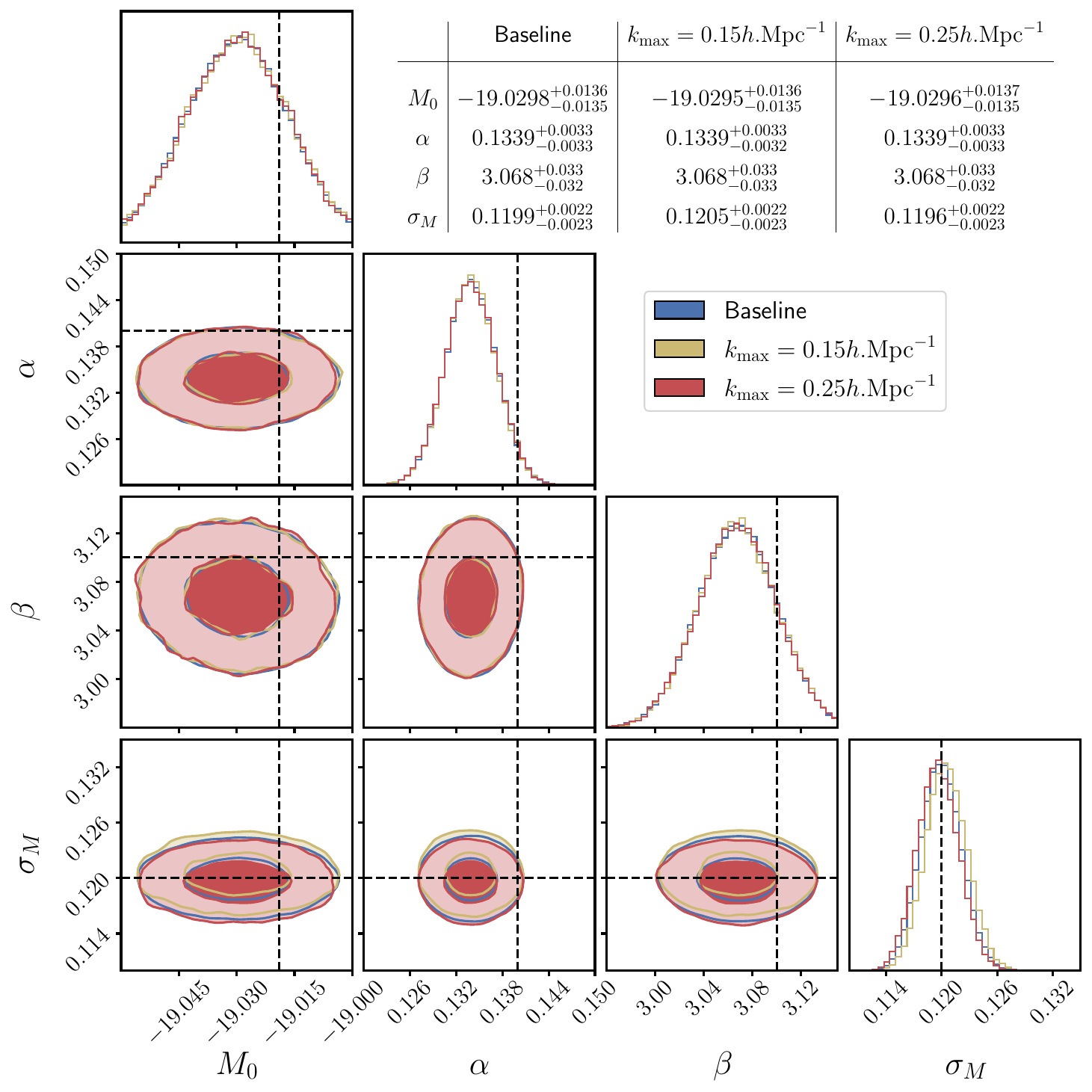}
    \caption{MCMC results of the Hubble diagram fit using three different values of $k_\mathrm{max}$ to compute the power spectrum $P_{\theta\theta}$ for the complete velocity covariance. The fit is performed on the same realization shown in Fig.~\ref{fig:mcmc_normvsjlavscov}. The dashed black lines show the simulation input value for each parameter. The baseline refers to $k_{max}=0.2$ \hmpc as described in Sect. \ref{Methodsec}. The table shows the median posterior probability, the fitted value, for each parameter and the relative error. }
    \label{fig:mcmc_diffkmax} 
\end{figure}

We perform the Hubble diagram fit on one random realization changing the maximum scale $k_\mathrm{max}$ in the velocity covariance computation. Figure~\ref{fig:mcmc_diffkmax} shows the result of this test. We find that changing $k_\mathrm{max}$ does not change the contours with respect to the baseline choice of $k_\mathrm{max}=0.2$ \hmpc\ done in the analysis.  As in the previous section, Fig.~\ref{fig:mcmc_diffkmax} shows that the recovered parameter values are exactly the same within the uncertainties. We conclude that for the purpose of systematic assessment we are independent from the choice of the maximum scale.

\section{Scatter of $\alpha$, $\beta$, and $\sigma_M$ across the 27 ZTF realizations}
\label{app:otherparams}
\begin{figure}
    \centering
    \includegraphics[width=\columnwidth]{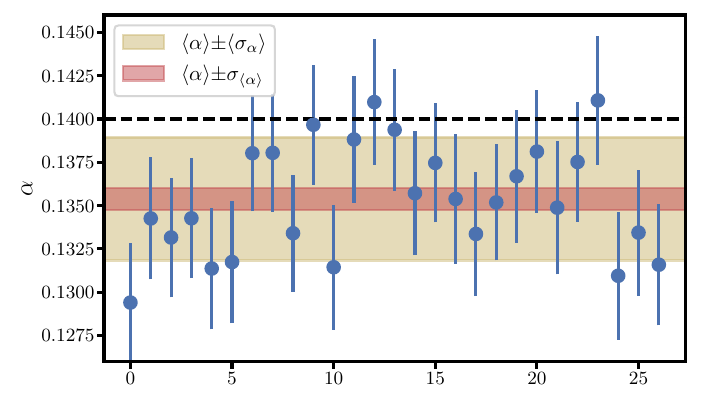}
    \includegraphics[width=\columnwidth]{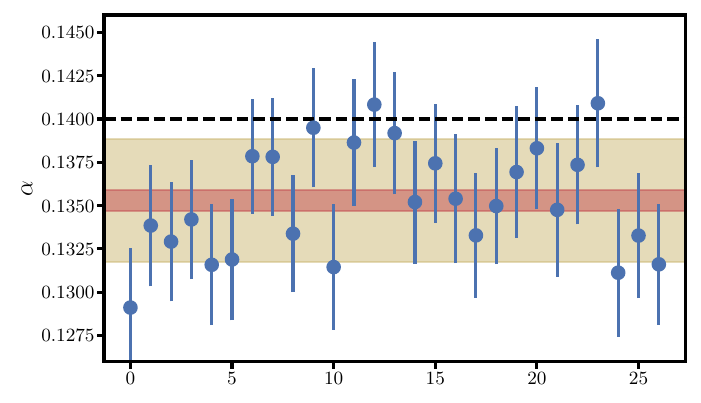}
    \includegraphics[width=\columnwidth]{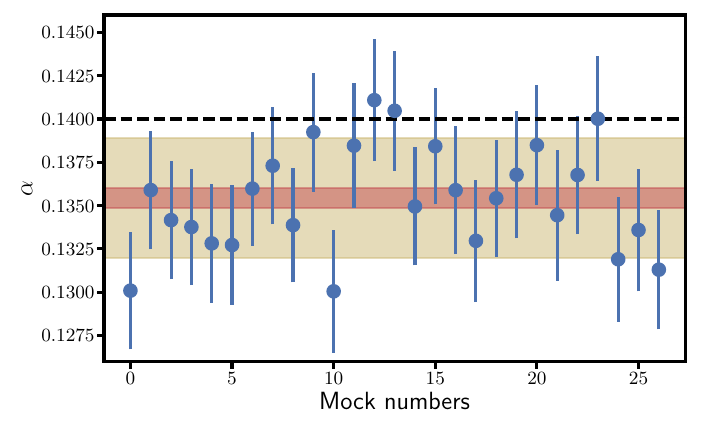}
    \caption{$\alpha$ values obtained from the fit of the 27 realizations. The dashed black lines show the simulation input value of the $\alpha$ parameter. \textit{Top panel}: Fit without velocity error term. \textit{Middle panel}: Fit with a velocity diagonal error term. \textit{Bottom panel}: Fit with a full velocity covariance. }
    \label{fig:simalpha}
\end{figure}

\begin{figure}
    \centering
    \includegraphics[width=\columnwidth]{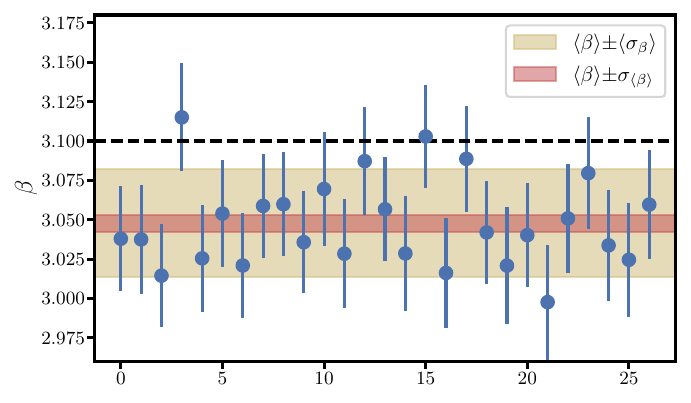}
    \includegraphics[width=\columnwidth]{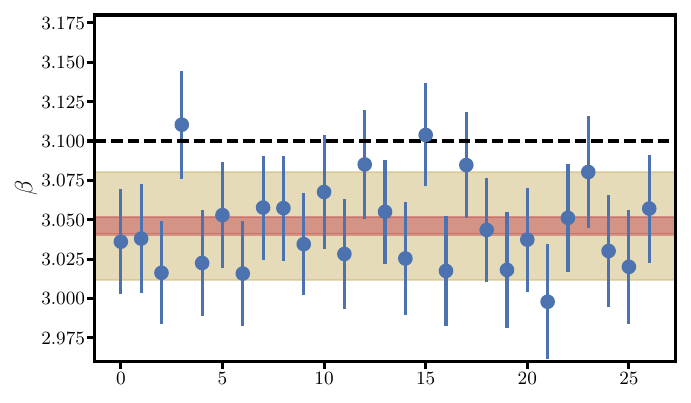}
    \includegraphics[width=\columnwidth]{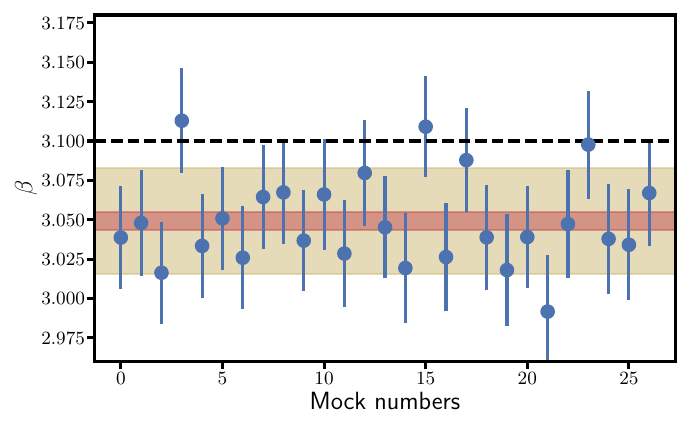}
    \caption{$\beta$ values obtained from the fit of the 27 realizations. The dashed black lines show the simulation input value of the $\beta$ parameter. \textit{Top panel}: Fit without velocity error term.  \textit{Middle panel}: Fit with a velocity diagonal error term. \textit{Bottom panel}: Fit with a full velocity covariance. }
    \label{fig:simbeta}
\end{figure}

In Sect.~\ref{simulationresults} we emphasized that using or not the PVs covariance does not affect the fit of the $\alpha$ and $\beta$ parameters. This is what we can see in more details in Figs.~\ref{fig:simalpha} and \ref{fig:simbeta}. The figures show the results for these parameters across the 27 ZTF realizations. On these figures we do not notice any significant changes. However, we see that the values of $\alpha$ and $\beta$ are slightly bias, as for $M_0$, this could be an effect of the sample selection. 

\begin{figure}
    \centering
    \includegraphics[width=\columnwidth]{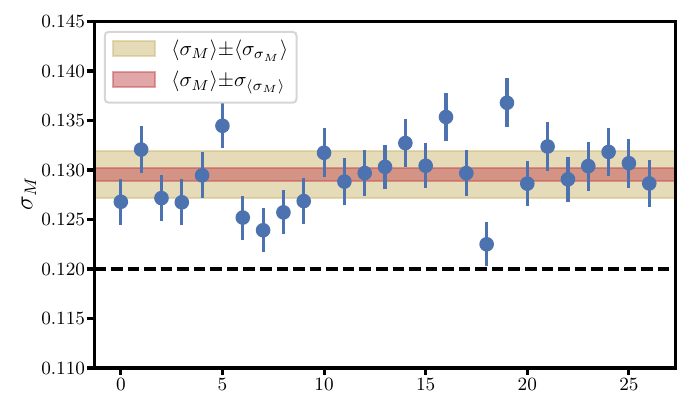}
    \includegraphics[width=\columnwidth]{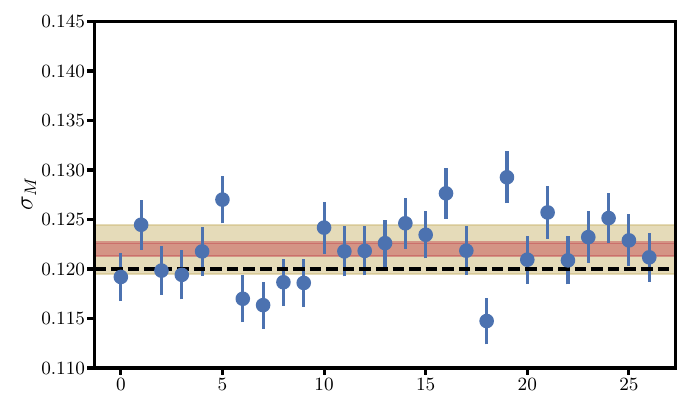}
    \includegraphics[width=\columnwidth]{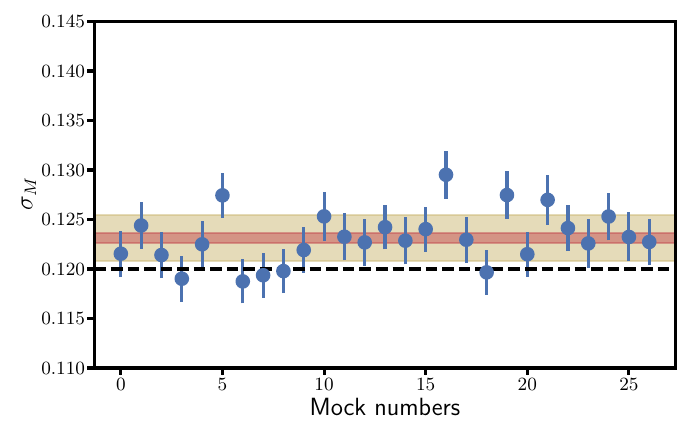}
    \caption{$\sigma_M$ values obtained from the fit of the 27 realizations. The dashed black lines show the simulation input value of the $\sigma_M$ parameter.\ \textit{Top panel}: Fit without velocity error term.  \textit{Mid panel}: Fit with a velocity diagonal error term.\ \textit{Bottom panel}: Fit with a full velocity covariance. }
    \label{fig:simsigM}
\end{figure}

Figure~\ref{fig:simsigM} shows the same results for the $\sigma_M$ parameter, which represents the non-modeled scattering of the Hubble residuals. We can see that including PVs as diagonal or full covariance matrix reduces the recovered value of $\sigma_M$. It can be explained by the fact that this parameter will absorb the PV scattering if it is not modeled.

\end{appendix}

\end{document}